\documentclass[prd,aps,twocolumn,nofootinbib,showpacs,floats,floatfix]{revtex4}
\usepackage{epsfig}
\usepackage{dcolumn}
\usepackage{bm}
\usepackage{color}
\usepackage{comment}
\usepackage{amsmath,amssymb}
\usepackage[utf8]{inputenc}
\DeclareGraphicsExtensions{.eps,.ps,.eps.gz,.ps.gz,.eps.Z,{}}

\graphicspath{{figures/}}

\newcommand{\bk}{{\bf k}}
\newcommand{\bx}{{\bf x}}
\newcommand{\bv}{{\bf v}}
\newcommand{\bn}{{\bf n}}
\newcommand{\bp}{{\bf p}}
\newcommand{\bq}{{\bf q}}
\newcommand{\br}{{\bf r}}
\newcommand{\bku}{\mathbf{k}\in {\rm uhs}}

\newcommand{\VV}{\mathcal{V}}

\newcommand{\HH}{{\cal H}}

\newcommand{\be}{\begin{equation}}
\newcommand{\ee}{\end{equation}}
\newcommand{\bea}{\begin{eqnarray}}
\newcommand{\eea}{\end{eqnarray}}

\begin{document}

\title{Probing redshift-space distortions with phase correlations}

\author{Felipe O. Franco$^{1}$, Camille Bonvin$^{1}$, Danail Obreschkow$^{2, 3}$, Kamran Ali$^{2, 3}$ and Joyce Byun$^{1}$}
\affiliation{${}^{1}$ D\'epartement de Physique Th\'eorique and Center for Astroparticle Physics (CAP), University of Geneva,
24 quai Ernest Ansermet, CH-1211 Geneva, Switzerland\\
${}^{2}$ International Centre for Radio Astronomy Research (ICRAR), University of Western Australia, 35 Stirling Highway,
Crawley WA 6009, Australia\\
${}^{3}$ ARC Centre of Excellence for All-sky Astrophysics (CAASTRO)}

\date{\today}

\begin{abstract}

Redshift-space distortions are a sensitive probe of the growth of large-scale structure. In the linear regime, redshift-space distortions are fully described by the multipoles of the two-point correlation function. In the nonlinear regime, however, higher-order statistics are needed to capture the full information of the galaxy density field. In this paper, we show that the redshift-space line correlation function—which is a measure of Fourier phase correlations—is sensitive to the nonlinear growth of the density and velocity fields and to the nonlinear mapping between real and redshift space. We expand the line correlation function in multipoles, and we show that almost all of the information is encoded in the monopole, quadrupole, and hexadecapole. We argue that these multipoles are highly complementary to the multipoles of the two-point correlation function: first, because they are directly sensitive to the difference between the density and the velocity coupling kernels, which is a purely nonlinear quantity; and second, because the multipoles are proportional to different combinations of  $f$ and $\sigma_8$. Measured in conjunction with the two-point correlation function and the bispectrum, the multipoles of the line correlation function could therefore allow us to disentangle efficiently these two quantities and to test modified theories of gravity.

\end{abstract}

\maketitle
 
\section{Introduction}
\label{sec:intro}

Cosmological galaxy redshift surveys, like the 6dF Galaxy Redshift Survey~\cite{Jones:2009yz}, Sloan Digital Sky Survey~\cite{Abazajian:2008wr}, WiggleZ survey~\cite{2012PhRvD..86j3518P}, VIPERS survey~\cite{2018A&A...609A..84S} or BOSS survey~\cite{2015ApJS..219...12A}, map the distribution of galaxies in redshift space. Since the redshift of galaxies is affected by their peculiar velocity, the observed galaxy distribution is slightly distorted with respect to the real-space galaxy distribution. In the linear regime, these redshift-space distortions modify the two-point correlation function and the power spectrum, by adding a quadrupole and an hexadecapole modulation in the signal~\cite{Kaiser:1987qv,Hamilton:1997zq}. Measuring these multipoles has been one of the main goals of recent redshift galaxy surveys; see, e.g.,~\cite{Alam:2016hwk}. These measurements have been very successful and have provided constraints on modified theories of gravity~\cite{Sanchez:2016sas}. Redshift-space distortions are indeed highly sensitive to the growth rate of perturbation $f$, which is generally modified in alternative theories of gravity.

In the nonlinear regime, the multipoles of the correlation function do however not fully trace the information present in galaxy surveys. The nonlinear gravitational evolution of the density and peculiar velocity generates indeed a flow of information into higher-order statistics. An obvious choice to capture this flow of information is to look at the three- point correlation function (or Fourier-space bispectrum); see, e.g.,~\cite{Scoccimarro:1999ed,Gaztanaga:2005ad,Gil-Marin:2014pva} and Refs. therein. However, this estimator is a three-dimensional function with significant redundancies with itself and the two-point statistics, making its computation and information analysis a complex task.

Various alternative observables have been constructed in order to access information in the nonlinear regime; see, e.g.,~\cite{Hikage:2002ki,Codis:2013exa,White:2016yhs,Scrimgeour:2012wt}. The goal of such observables is twofold: first, part of the information present in the bispectrum has already been measured in the power spectrum. One can then wonder if it is possible to construct an observable which is less redundant with the power spectrum. And second, since the bispectrum is complicated in redshift space, it would be interesting to construct an estimator which encodes the same type of information, but which is simpler to model.

In this paper, we study one possible alternative: the line correlation function. The line correlation function has been introduced in~\cite{Obreschkow:2012yb} and analytically modeled in real space in~\cite{Wolstenhulme:2014cla}. This observable is constructed from correlations between the {\it phases} of the density field. Since the two-point function is only sensitive to the {\it amplitude} of the density field, it seems promising to use in conjunction an observable which is targeted to measure the phases (see~\cite{Scherrer:1990dq,Ryden:1991fi,Soda:1991jz,Jain:1998it,Chiang:1999hm,Coles:2000zg,Chiang:2000vt,Watts:2002py,Chiang:2002vm,Chiang:2002sb,Coles:2003dw,Matsubara:2003te,Hikage:2003kr,Hikage:2005ia,Szepietowski:2013ura} for other observables based on phase correlations). Fisher forecasts in real space have shown that combining the line correlation function with the two-point correlation function does indeed improve parameter constraints on $\Lambda$CDM cosmology by up to a factor of 2~\cite{Eggemeier:2016asq,Byun:2017fkz}. The gain obtained from the line correlation function in the case of a warm dark matter model or alternative theories of gravity, like the symmetron and $f(R)$ model is even stronger~\cite{Ali:2018sdk}.

Here we derive an expression for the line correlation function in redshift space. We show that the line correlation function can be expanded in Legendre polynomials and that almost all of the information is encoded in the first even three multipoles, i.e., the monopole, quadrupole, and hexadecapole, similarly to the two-point correlation func- tion. These multipoles are sensitive to the nonlinear coupling kernels of the density and of the peculiar velocity. As such, the line correlation function provides a simple way to probe the nonlinear evolution in redshift space and consequently, to constrain alternative theories of gravity in the nonlinear regime, for example, at the scales where screening mechanisms start to act. Note that our approach differs and complements the work of~\cite{Eggemeier:2015ifa}, which studies a modified version of the line correlation function (using an anisotropic window function) and is targeted to measure an anisotropic signal in two-dimensional Zel’dovich mock density fields.

Let us mention that throughout this paper we will use second-order perturbation theory to model redshift-space distortions. Studies of the bispectrum have shown that this approach does not fully account for nonlinearities in redshift space. Various models have been developed over the years to provide a more accurate description of redshift- space distortions, either by improving on perturbation theory~\cite{Scoccimarro:1999ed,Taruya:2010mx,Hashimoto:2017klo}, or by building effective coupling kernels based on numerical simulations~\cite{Gil-Marin:2014pva}. In particular, these models are able to describe the Fingers of God, which are not accounted for at second order in perturbation theory. However, since it is not clear which of these models is more adapted to describe phase correlations, we start by using only second-order perturbation theory. Our modeling should therefore be regarded as a first step towards an accurate description of phase correlations in redshift space. In a future work, we will compare our modeling with measurements of the line correlation function in numerical simulations, in order to improve the description of redshift-space distortions in the strongly nonlinear regime.

The remainder of the paper is organized as follows. In Sec.~\ref{sec:line} we derive an expression for the line correlation function in redshift space, at second order in perturbation theory. In Sec.~\ref{sec:mult} we expand the line correlation function in Legendre polynomials. We derive a general expression valid for any multipole $n$. In Sec.~\ref{sec:result} we calculate numerically the first multipoles in a $\Lambda$CDM universe and we show that the multipoles larger than $n=4$ are negligible. We conclude in Sec.~\ref{sec:conclusion}.

\section{The line correlation function of the observed number counts}
\label{sec:line}

Galaxy surveys measure the overdensity of galaxies in redshift space,
\be
\Delta(\bn, z)=\frac{N(\bn, z)-\bar N(z)}{\bar N(z)}\, ,
\ee
where $N(\bn, z)$ denotes the number of galaxies detected in a pixel situated at redshift $z$ and in direction $\bn$, and $\bar N(z)$ is the average number of galaxies per pixel at a given redshift. The Fourier transform of the galaxy overdensity\footnote{We use the Fourier convention $f\left(\mathbf{x}\right)=\int d^{3}\mathbf{k}e^{i\mathbf{x}\cdot\mathbf{k}}f\left(\mathbf{k}\right)$ and $f\left(\mathbf{k}\right)=\int\frac{d^{3}\mathbf{x}}{\left(2\pi\right)^{3}}e^{-i\mathbf{x}\cdot\mathbf{k}}f\left(\mathbf{x}\right)\,.$}, $\Delta(\bk, z)$, is characterised by an amplitude $|\Delta(\bk, z)|$ and a phase
\be
\epsilon_\Delta(\bk, z)\equiv\frac{\Delta(\bk, z)}{|\Delta(\bk, z)|}\, .
\ee
The line correlation function of $\Delta$ is then defined as 
\begin{align}
&\ell\left(\mathbf{r},z\right)=\frac{V^{3}}{\left(2\pi\right)^{9}}\!\left(\frac{r^{3}}{V}\right)^{3/2}\hspace{-0.4cm}\langle\epsilon_\Delta\left(\mathbf{s}, z\right)\epsilon_\Delta\left(\mathbf{s}+\mathbf{r}, z\right)\epsilon_\Delta\left(\mathbf{s}-\mathbf{r}, z\right)\rangle\nonumber\\
&=\frac{V^{3}}{\left(2\pi\right)^{9}}\left(\frac{r^{3}}{V}\right)^{3/2}\hspace{-0.2cm}\underset{\substack{|\bk_1|, |\bk_2|,\\ |\bk_3| \leq 2\pi/r}}{\int\!\!\int\!\!\int}d^{3}\mathbf{k}_{1}d^{3}\mathbf{k}_{2}d^{3}\mathbf{k}_{3}e^{i\left(\mathbf{k}_{1}+\mathbf{k}_{2}+\mathbf{k}_{3}\right)\cdot\mathbf{s}}\nonumber\\
&\quad\times e^{i\left(\mathbf{k}_{2}-\mathbf{k}_{3}\right)\cdot\mathbf{r}}\langle\epsilon_\Delta(\mathbf{k}_{1},z)\epsilon_\Delta(\mathbf{k}_{2}, z)\epsilon_\Delta(\mathbf{k}_{3}, z)\rangle\,,\label{linedef}
\end{align}
where $\epsilon_\Delta(\mathbf{s}, z)$ is the inverse Fourier transform of $\epsilon_\Delta(\bk, z)$. As discussed in~\cite{Obreschkow:2012yb}, the cutoff at high $k$ has been introduced to avoid the divergence of the line correlation function due to an infinite number of phase factors at arbitrarily small scales, which do not carry any information.

We start by calculating the three-point correlation function of the phase of $\Delta(\bk, z)$. For this, we need a description of the galaxy number count valid at second order in perturbation theory. At first order in perturbation theory, we have the standard Kaiser expression,
\be
\label{Delta1}
\Delta^{(1)}(\bn, z)=b_1\delta^{(1)}-\frac{1}{\HH}\partial_\chi (\bv^{(1)}\!\cdot\bn)\, ,
\ee
where $\HH=(da/d\eta)/a$ is the Hubble parameter in conformal time $\eta$, $b_1$ is the linear bias, $\delta$ is the matter density field, $\bv$ is the peculiar velocity of galaxy and $\partial_\chi$ denotes radial derivative ($\chi$ being the conformal distance). The second term in Eq.~\eqref{Delta1} represents the contribution from redshift-space distortions. Note that $\Delta$ contains various other contributions, namely relativistic effects and lensing effects~\cite{Yoo:2009au,Bonvin:2011bg,Challinor:2011bk}, but we neglect these terms here since we are mainly interested in small scales and low redshifts, where they are expected to be subdominant.

At second order in perturbation theory, we can identify two types of contributions: first, the contribution coming from the nonlinear gravitational evolution of the density and peculiar velocity field. We call this contribution the {\it intrinsic} contribution, $\Delta^{(2)}_{\rm int}$, because it is due to the fact that the density, velocity, and bias are intrinsically nonlinear quantities,
\begin{align}
\label{Deltaint}
&\Delta^{(2)}_{\rm int}(\bn, z)=b_1\delta^{(2)}-\frac{1}{\HH}\partial_\chi (\bv^{(2)}\!\cdot\bn)\\
&+\frac{b_2}{2}\left[ \big(\delta^{(1)}\big)^2-\big\langle\big(\delta^{(1)}\big)^2 \big\rangle\right]
+\frac{b_{s^2}}{2}\left[ s^2-\big\langle s^2 \big\rangle\right]\, ,\nonumber
\end{align}
where $s$ denotes the tidal tensor~\cite{McDonald:2009dh}.

In addition, at second order, we have contributions coming from the fact that the mapping between real space and redshift space is itself nonlinear. We call these contributions $\Delta^{(2)}_{\rm map}$. A detailed derivation is given in Appendix~\ref{app:Delta}, following~\cite{Nielsen:2016ldx}. The result is
\begin{align}
\Delta^{(2)}_{\rm map}(\bn, z)=&-\frac{b_1}{\HH}\delta^{(1)}\partial_\chi (\bv^{(1)}\!\cdot\bn)-\frac{b_1}{\HH}\partial_\chi\delta^{(1)}(\bv^{(1)}\!\cdot\bn)\nonumber\\
&+\frac{1}{\HH^2}\partial_\chi\Big[\partial_{\chi}(\bv^{(1)}\!\cdot\bn)(\bv^{(1)}\!\cdot\bn)\Big]\, . \label{eq:Deltamap}
\end{align}
Note that besides these dominant contributions, many other terms contribute to $\Delta_{\rm map}$ at second order, due to lensing and relativistic effects~\cite{Bertacca:2014dra,Yoo:2014sfa,DiDio:2014lka}. But again we neglect them here since they are expected to become relevant on larger scales and higher redshifts.

Since $\Delta$ is expressed in terms of the observed coordinates $\mathbf{x}_{\rm obs}=\chi(z)\bn$, where $\chi$ is the comoving distance evaluated at the observed redshift $z$, we can consistently Fourier transform it,
\be
\Delta(\bk, z)=\frac{1}{(2\pi)^3}\int d^3\bx_{\rm obs}\,e^{-i\bk\cdot\bx_{\rm obs}}\Delta(\bx_{\rm obs},z)\, .
\ee
At first order in perturbation theory, we obtain
\begin{align}
\Delta^{(1)}(\bk, z)= &b_1\,\delta^{(1)}(\bk, z)-\frac{1}{\HH}(\hat \bk\cdot\bn)^2 V^{(1)}(\bk, z)\, ,\nonumber\\
=&\Big(b_1+f\mu^2 \Big)\delta^{(1)}(\bk, z)\, ,\label{Deltak1}
\end{align} 
where $V$ is related to the Fourier transform of $\bv$ by
\begin{align}
\bv(\bk, z)=-i\frac{\hat \bk}{k}V(\bk, z)\, ,
\end{align}
and $\mu=\hat{\bk}\cdot \bn$.
In the second line of~\eqref{Deltak1}, we have used the continuity equation at linear order to relate the velocity to the density, and we define the growth rate as
\be
f=\frac{d\ln D_1}{d\ln a}\, ,
\ee
where $D_1$ is the linear growth function. 

At second order in perturbation theory, the density field $\delta^{(2)}$ and the velocity field $V^{(2)}$ take the form,
\begin{align}
\delta^{(2)}(\bk, z)=&\int d^3\bq_1 \int d^3\bq_2\,\delta_D(\bk-\bq_1-\bq_2)\nonumber\\
&F_2(\bq_1,\bq_2) \delta^{(1)}(\bq_1, z)\delta^{(1)}(\bq_2, z)\, ,\\
V^{(2)}(\bk, z)=&-\HH(z) f(z) \int d^3\bq_1 \int d^3\bq_2\,\delta_D(\bk-\bq_1-\bq_2)\nonumber\\
&G_2(\bq_1,\bq_2) \delta^{(1)}(\bq_1, z)\delta^{(1)}(\bq_2, z)\, ,
\end{align}
where the nonlinear kernels are given by~\cite{Bernardeau:2001qr,Bernardeau:2011sf}
\begin{align}
F_{2}\left(\mathbf{k}_{1},\mathbf{k}_{2}\right)  =&  \frac{1+\epsilon_{F}}{2}+\frac{\widehat{\mathbf{k}}_{1}\cdot\widehat{\mathbf{k}}_{2}}{2}\left(\frac{k_{1}}{k_{2}}+\frac{k_{2}}{k_{1}}\right)\\
&+\frac{1-\epsilon_{F}}{2}\left(\widehat{\mathbf{k}}_{1}\cdot\widehat{\mathbf{k}}_{2}\right)^{2}\,,\nonumber\\
G_{2}\left(\mathbf{k}_{1},\mathbf{k}_{2}\right) = & \epsilon_{G}+\frac{\widehat{\mathbf{k}}_{1}\cdot\widehat{\mathbf{k}}_{2}}{2}\left(\frac{k_{1}}{k_{2}}+\frac{k_{2}}{k_{1}}\right)\\
&+\left(1-\epsilon_{G}\right)\left(\widehat{\mathbf{k}}_{1}\cdot\widehat{\mathbf{k}}_{2}\right)^{2}\,,\nonumber
\end{align}
with $\epsilon_{F}\simeq\left(3/7\right)\Omega_{m}(z)^{-1/143}$ and $\epsilon_{G}=\epsilon_{F}+\left(3/2\right)\left(\epsilon_{F}-3/7\right)$. The kernels depend therefore very mildly on $z$ through $\epsilon_F$ and $\epsilon_G$.

With this, the intrinsic part becomes at second order
\begin{align}
&\Delta_{\rm int}^{(2)}(\bk, z)=\int d^3\bq_1 \int d^3\bq_2\,\delta_D(\bk-\bq_1-\bq_2)\nonumber\\
&\left[b_1 F_2(\bq_1,\bq_2)+\frac{b_2}{2}+\frac{b_{s^2}}{2}S_2(\bq_1,\bq_2)+f \mu^2G_2(\bq_1,\bq_2)\right]\nonumber\\
&\times\delta^{(1)}(\bq_1, z)\delta^{(1)}(\bq_2, z)\, ,
\end{align}
where $\mu=\hat{\bk}\cdot\bn$ and $S_2(\bq_1,\bq_2)=\left(\widehat{\bq}_1\cdot \widehat{\bq}_2\right)^2-1/3$.

Finally, the mapping part at second order can be written as
\begin{align}
\Delta_{\rm map}^{(2)}(\bk, z)&=\int d^3\bq_1 \int d^3\bq_2\,\delta_D(\bk-\bq_1-\bq_2)\nonumber\\
&\frac{f\mu k}{2}\left[\frac{\mu_1}{q_1}\Big(b_1+f\mu_2^2 \Big)+\frac{\mu_2}{q_2}\Big(b_1+f\mu_1^2 \Big) \right]\nonumber\\
&\times\delta^{(1)}(\bq_1, z)\delta^{(1)}(\bq_2, z)\, , \label{eq:Deltamapk}
\end{align}
where $\mu_i=\hat{\bq}_i\cdot\bn, \ i=1,2$. Note that the mapping term is often computed in a different way. Instead of writing $\Delta$ as a function of the observed coordinate $\bx_{\rm obs}$ as we did here and then Fourier transform it, one can express $\Delta$ in terms of the unperturbed coordinate $\bx$ and then expand the exponential in the Fourier transform around $\bx$ (see Appendix~\ref{app:Delta} for more detail). This procedure gives rise to the same expression for $\Delta_{\rm map}^{(2)}(\bk, z)$. Note that this term is sometimes defined as part map of the Finger of God contribution, since it arises from the exponential in the Fourier transform. We reserve however this name for the damping due to the random motion of galaxies at small scales (see, e.g., ~\cite{Scoccimarro:1999ed,Song:2015gca}), which we do not include in our derivation, since it is not captured by second-order perturbation theory.

We now compute the three-point correlation function of the phase of $\Delta(\bk, z)$, which enters in Eq.~\eqref{linedef}. We have
\be
\label{3eps}
\langle \epsilon_\Delta(\bk_1)\epsilon_\Delta(\bk_2)\epsilon_\Delta(\bk_3)\rangle=\int [d\theta]\mathcal{P}[\theta]\epsilon_\Delta(\bk_1)\epsilon_\Delta(\bk_2)\epsilon_\Delta(\bk_3)\, ,
\ee
where $\mathcal{P}[\theta]$ is the probability distribution function of the field $\theta(\bk)$ defined through $\epsilon_\Delta(\bk)=e^{i\theta(\bk)}$. Note that here we have dropped the dependence in redshift $z$ in the argument of $\epsilon_\Delta$ to ease the notation. Following~\cite{Matsubara:2003te,Wolstenhulme:2014cla}, we start by expressing the probability distribution function for $\Delta(\bk, z)$ using the Edgeworth expansion~\cite{1991ApJ...381..349S,1995ApJ...442...39J,1995ApJ...443..479B}, which is valid for mildly non-Gaussian fields,
\begin{align}
\label{P_Edg}
&\mathcal{P}[\Delta] =N_G \exp \left(-\frac{1}{2}\int d^3\bk \frac{\Delta(\bk) \Delta(-\bk)}{P_\Delta(\bk)} \right)\Bigg\{1\,+\\
& \frac{1}{3!}\! \int\!\! d^3\bp\, d^3\bq \, \frac{B_\Delta(\bp,\bq, -\bp-\bq)\Delta(-\bp)\Delta(-\bq)\Delta(\bp+\bq)}{P_\Delta(\bp)P_\Delta(\bq)P_\Delta(\bp+\bq)} \Bigg\} \nonumber
\end{align}
where $N_G$ is a normalisation factor. Here, $P_\Delta(\bk)$ and $B_\Delta(\bp,\bq,\bk)$ are the power spectrum and bispectrum of $\Delta$ defined through
\bea
\langle \Delta(\bk)\Delta(\bk')\rangle&=&P_\Delta(\bk)\delta_D(\bk+\bk')\, ,\\
\langle \Delta(\bp)\Delta(\bq)\Delta(\bk)\rangle&=&B_\Delta(\bp,\bq)\delta_D(\bp+\bq+\bk)\, .
\eea
Note that since redshift-space distortions break statistical isotropy, $P_\Delta(\bk)$ depends not only on the modulus of $\bk$ but also on its orientation with respect to the direction of observation $\bn$. Similarly, the bispectrum depends not only on the shape of the triangle but also on its orientation.

Following the derivation in~\cite{Wolstenhulme:2014cla}, we first discretize the field $\Delta(\bk)\rightarrow\Delta_\bk$ for a finite survey volume, and then we integrate over the amplitude $|\Delta_\bk|$ to obtain the probability distribution function of the phase,
\begin{align}
&\mathcal{P}(\{\theta_\bk \})\!\!\!\prod_{\bku} \!\!\!d\theta_\bk=\!\Bigg\{1+\frac{\sqrt{\pi}}{6}\!\!\!\sum_{\bp\in{\rm uhs}}\!\!
b_\Delta(\bp,\bp)\cos(2\theta_{\bp}\!-\!\!\theta_{2\bp}\!)\nonumber\\
&+\frac{1}{3}\left(\frac{\sqrt{\pi}}{2} \right)^3\!\!\!\sum_{\bp\neq \bq\in {\rm uhs}}
 \!\!\!\!\Big[b_\Delta(\bp,\bq)\cos(\theta_{\bp}+\theta_{\bq}-\theta_{\bp+\bq})\nonumber\\
&+ b_\Delta(\bp,-\bq)\cos(\theta_{\bp}-\theta_{\bq}+\theta_{\bp-\bq})\Big] \Bigg\}
\prod_{\bku} \frac{d\theta_\bk}{2\pi}\, , \label{Probability}
\end{align}
where we have defined
\be
\label{defb}
b_\Delta(\bp,\bq)\equiv\sqrt{\frac{(2\pi)^3}{V}}\frac{B_\Delta(\bp,\bq)}{\sqrt{P_\Delta(\bp)P_\Delta(\bq)P_\Delta(\bk)}}\, .
\ee
Inserting Eq.~\eqref{Probability} into~\eqref{3eps}, we obtain in the continuous limit,
\begin{align}
\label{eps3}
\langle \epsilon_\Delta(\bk_1)&\epsilon_\Delta(\bk_2)\epsilon_\Delta(\bk_3)\rangle=\\
&\frac{(2\pi)^3}{V}\left(\frac{\sqrt{\pi}}{2}\right)^3b_\Delta(\bk_1,\bk_2,\bk_3)\,\delta_D(\bk_1+\bk_2+\bk_3)\, .\nonumber
\end{align}
At second order in perturbation theory, we have
\begin{align}
&b_\Delta(\bk_1,\bk_2,\bk_3)=2\sqrt{\frac{\left(2\pi\right)^{3}}{V}}
\label{b_fin}\\
&\times\left[W_{2}\left(\mathbf{k}_{1},\mathbf{k}_{2},\mathbf{k}_{3},\bn\right)\sqrt{\frac{P_{L}\left(k_{1}, z\right)P_{L}\left(k_{2},z\right)}{P_{L}\left(k_{3},z\right)}}+\text{cyc}\right]\,  ,\nonumber
\end{align}
where $P_L$ denotes the linear power spectrum of $\delta$ at redshift $z$ and
\begin{align}
W_{2}\left(\mathbf{k}_{1},\mathbf{k}_{2},\mathbf{k}_{3},\bn\right)=&W^{\rm int}_{2}\left(\mathbf{k}_{1},\mathbf{k}_{2},\mathbf{k}_{3},\bn\right)\label{eq:W2sum}\\
&+W^{\rm map}_{2}\left(\mathbf{k}_{1},\mathbf{k}_{2},\mathbf{k}_{3},\bn\right)\,.\nonumber
\end{align}
The intrinsic and mapping kernels read
\begin{align}
&W^{\rm int}_{2}=\frac{1}{b_1+\mu_3^2f}\times\label{eq:W2int}\\
& \left[b_1F_{2}\left(\mathbf{k}_{1},\mathbf{k}_{2}\right)+\frac{b_2}{2}+\frac{b_{s^2}}{2}S_2\left(\mathbf{k}_{1},\mathbf{k}_{2}\right)+\mu_3^2f\,G_{2}\left(\mathbf{k}_{1},\mathbf{k}_{2}\right)\right]\,,\nonumber\\
&W^{\rm map}_{2}=-\frac{f\mu_3k_3\Big[\frac{\mu_1}{k_1}(b_1+f\mu_2^2)+\frac{\mu_2}{k_2}(b_1+f\mu_1^2)  \Big]}{2\big(b_1+\mu_3^2f\big)}\,,\label{eq:W2map}
\end{align}
with $\mu_i=\widehat{\bk}_i\cdot\bn$ for $i=1,2,3$.

We see that the phase correlation of the observed number count $\Delta$ is sensitive to the nonlinear coupling kernel of the density field $F_2$, to the nonlinear coupling kernel of the velocity field $G_2$, and to the growth rate $f$. Since redshift-space distortions are not isotropic, the phase correlations depend on the direction of observation~$\bn$. Note that here we work in the distant-observer approximation, where $\bn$ is the same for all galaxies. 

The line correlation function is obtained by inserting~\eqref{b_fin} and~\eqref{eps3} into~\eqref{linedef}. We get
\begin{align}
&\ell\left(\mathbf{r},z\right)=\frac{r^{9/2}}{8\sqrt{2}\left(2\pi\right)^{3}}\hspace{-0.2cm}\underset{\substack{|\bk_1|, |\bk_2|,\\ |\bk_1 + \bk_2| \leq 2\pi/r}}{\int\!\!\int}\hspace{-0.4cm}d^{3}\mathbf{k}_{1}d^{3}\mathbf{k}_{2}\label{line_beta}\\
& W_{2}\left(-\mathbf{k}_{1}-\mathbf{k}_{2},\mathbf{k}_{1},\mathbf{k}_{2},\bn\right)\sqrt{\frac{P_{L}\left(\left|\mathbf{k}_{1}+\mathbf{k}_{2}\right|,z\right)P_{L}\left(k_{1},z\right)}{P_{L}\left(k_{2},z\right)}}\nonumber \\
 &\left[e^{i\left(\mathbf{k}_{1}-\mathbf{k}_{2}\right)\cdot\mathbf{r}}+e^{i\left(\mathbf{k}_{1}+2\mathbf{k}_{2}\right)\cdot\mathbf{r}}+e^{-i\left(2\mathbf{k}_{1}+\mathbf{k}_{2}\right)\cdot\mathbf{r}}\right]\, .\nonumber
\end{align}
Here, we have used the Dirac Delta function to rewrite the three permutations in~\eqref{b_fin} with the same kernel $W_2$ multiplied by three different exponentials. In this way, the kernel $W_2$ depends on the direction of observation $\bn$ only through its scalar product with $\bk_2$. We will see that this property is useful to solve analytically some of the integrals in~\eqref{line_beta}. 

\begin{figure}[t]
\centering
\includegraphics[width=0.35\columnwidth]{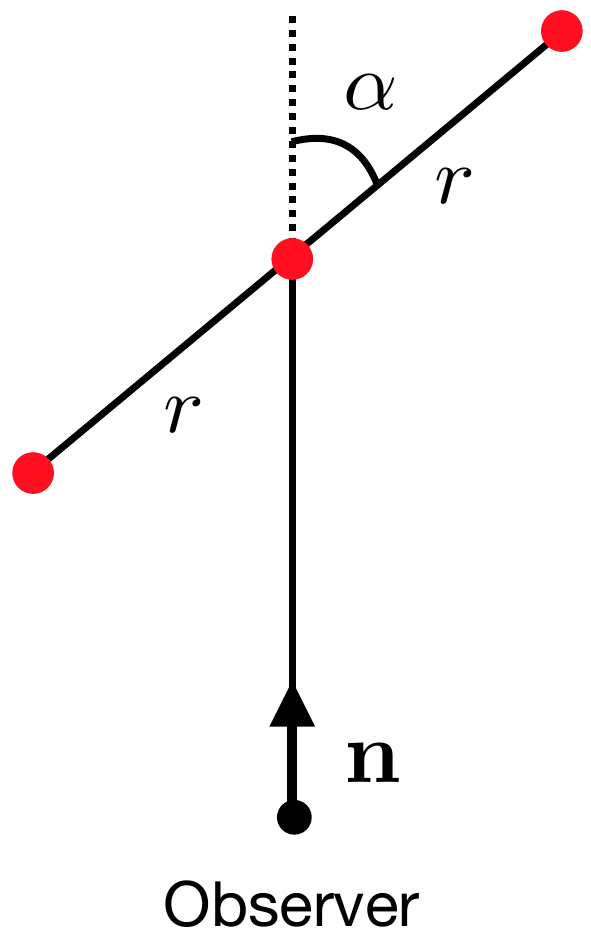}
\caption{\label{fig:coordinate} Representation of the coordinate system used to express the line correlation function: $r=|\mathbf{r}|$ denotes the separation between galaxies, and $\cos\alpha=\hat\br\cdot\bn$ is the orientation of the line with respect to the direction of observation $\bn$. We work in the distant-observer approximation, in which $\bn$ is the same for all galaxies.}
\end{figure}

Since redshift-space distortions break isotropy, the line correlation function depends not only on the modulus of the separation $r=|\mathbf{r}|$, but also on the orientation of the vector $\br$ with respect to the line-of-sight: $\cos\alpha=\hat\br\cdot\bn$, as depicted in Fig.~\ref{fig:coordinate}.
In the rest of this paper, we will study the dependence of the line correlation function on $\alpha$. Note that in the case where $f=0$, Eq.~\eqref{line_beta} is equivalent to the expression derived in~\cite{Wolstenhulme:2014cla}.

\section{Multipole expansion of the line correlation function}
\label{sec:mult}

In redshift space, the two-point correlation function of~$\Delta$ can be written as the sum of a monopole, quadrupole, and hexadecapole in the angle $\alpha$. At linear order in perturbation theory and using the distant-observer approximation, one can show that these three multipoles encode all the infor- mation present in the two-point correlation function~\cite{Hamilton:1997zq}. 

Contrary to the two-point correlation function, the line correlation function cannot be simply expressed as a sum of the first three even Legendre polynomials only. However, we will see that the contribution from the multipoles larger than $n=4$ is actually negligible so that most of the information about redshift-space distortions is indeed encoded in the monopole, quadrupole, and hexadecapole of $\ell$.

Since the Legendre polynomials form a basis, we can expand the line correlation function as
\be
\label{mult_exp}
\ell(r,\alpha, z)=\sum_{n=0}^{\infty}Q_n(r, z)L_n(\cos\alpha)\, ,
\ee
where $\cos\alpha=\hat\br\cdot\bn$ and $L_n$ denotes the Legendre polynomial of order $n$. The multipole of order $n$ can be measured by weighting the line correlation function by the appropriate Legendre polynomial
\be
Q_n(r, z)=\frac{2n+1}{2}\int_{-1}^1d\mu\, \ell(r,\mu, z)L_n(\mu)\, ,
\ee
where $\mu=\cos\alpha$.

To calculate explicitly $Q_n$, we insert Eq.~\eqref{line_beta} into~\eqref{mult_exp} and we expand the exponentials in~\eqref{line_beta} and the Legendre polynomial in~\eqref{mult_exp} in terms of spherical harmonics
\bea
e^{i \bk\cdot\br}&=&4\pi\sum_{n=0}^{\infty}\sum_{m=-n}^{n}i^{n}j_{n}\left(kr\right)Y_{n m}^{*}(\hat{\mathbf{k}})Y_{n m}(\hat{\mathbf{r}})\,,\\
L_n(\mu)&=&\frac{4\pi}{2n+1}\sum_{m=-n}^{n}Y_{n m}(\bn)Y_{n m}^*(\hat\br)\, .
\eea
We obtain
\begin{align}
&Q_{n}\left(r,z\right) = \frac{r^{9/2}}{8\sqrt{2}\left(2\pi\right)^{3}}8\pi^2\sum_{m=-n}^{n}\sum_{n^{\prime}=0}^{\infty}\sum_{m^{\prime}=-n^{\prime}}^{n^{\prime}}i^{n^{\prime}}\nonumber\\
&\int_{-1}^{1}d\mu\, Y_{n m}^{*}(\hat{\mathbf{r}})Y_{n^{\prime}m^{\prime}}(\hat{\mathbf{r}})Y_{n m}\left(\mathbf{n}\right) \hspace{-0.2cm}\underset{\substack{|\bk_1|, |\bk_2|,\\ |\bk_1 + \bk_2| \leq 2\pi/r}}{\int\!\!\int}\hspace{-0.4cm}d^{3}\mathbf{k}_{1}d^{3}\mathbf{k}_{2}\nonumber \\
 &W_{2}\left(-\mathbf{k}_{1}-\mathbf{k}_{2},\mathbf{k}_{1},\mathbf{k}_{2},\bn\right)\sqrt{\frac{P_{L}\left(\left|\mathbf{k}_{1}+\mathbf{k}_{2}\right|,z\right)P_{L}\left(k_{1},z\right)}{P_{L}\left(k_{2},z\right)}}\nonumber \\
 & \times\Big[j_{n^{\prime}}\left(\kappa_{1}r\right)Y_{n^{\prime}m^{\prime}}^{*}\left(\widehat{\boldsymbol{\kappa}}_{1}\right)+j_{n'}\left(\kappa_{2}r\right)Y_{n^{\prime}m^{\prime}}^{*}\left(\widehat{\boldsymbol{\kappa}}_{2}\right)\nonumber\\
 &\hspace{0.8cm}+j_{n^{\prime}}\left(\kappa_{3}r\right)Y_{n^{\prime}m^{\prime}}^{*}\left(\widehat{\boldsymbol{\kappa}}_{3}\right)\Big]\, ,\label{QellY}
 \end{align}
 where 
\bea
\boldsymbol{\kappa}_{1}&\equiv&\mathbf{k}_{1}-\mathbf{k}_{2}\,,\label{kappa1}\\
\boldsymbol{\kappa}_{2}&\equiv&\mathbf{k}_{1}+2\mathbf{k}_{2}\,,\label{kappa2}\\
\boldsymbol{\kappa}_{3}&\equiv&-2\mathbf{k}_{1}-\mathbf{k}_{2}\, .\label{kappa3}
\eea
Since in the distant-observer approximation, the direction of observation $\bn$ is fixed for all galaxies, we can choose $\bn$ on the $\hat{\mathbf{z}}$ axis without loss of generality. The integral over $\mu$ in Eq.~\eqref{QellY} becomes then an integral over the direction of $\br$, which can be performed and gives rise to $\delta_{nn'}\delta_{mm'}$. Combining the remaining spherical harmonics into Legendre polynomials, we obtain
\begin{align}
&Q_{n}\left(r,z\right)=\frac{r^{9/2}\left(2n+1\right)}{8\sqrt{2}\left(2\pi\right)^{3}}i^{n}\underset{\substack{|\bk_1|, |\bk_2|,\\ |\bk_1 + \bk_2| \leq 2\pi/r}}{\int\!\!\int}\hspace{-0.4cm}d^{3}\mathbf{k}_{1}d^{3}\mathbf{k}_{2}\nonumber \\
 &  W_{2}\left(-\mathbf{k}_{1}-\mathbf{k}_{2},\mathbf{k}_{1},\mathbf{k}_{2}, \bn\right)\sqrt{\frac{P_{L}\left(\left|\mathbf{k}_{1}+\mathbf{k}_{2}\right|,z\right)P_{L}\left(k_{1},z\right)}{P_{L}\left(k_{2},z\right)}}\nonumber\\
 &\times\Big[j_{n}\left(\kappa_{1}r\right)L_{n}\left(\widehat{\boldsymbol{\kappa}}_{1}\cdot\mathbf{n}\right)+
 j_{n}\left(\kappa_{2}r\right)L_{n}\left(\widehat{\boldsymbol{\kappa}}_{2}\cdot\mathbf{n}\right)\nonumber\\
 &\hspace{0.9cm}+j_{n}\left(\kappa_{3}r\right)L_{n}\left(\widehat{\boldsymbol{\kappa}}_{3}\cdot\mathbf{n}\right)\Big]\,.\label{QellP}
\end{align}
Equation~\eqref{QellP} contains a six-dimensional integral. We now show how to reduce it to a three-dimensional integral that we can compute numerically. 

\begin{figure}[t]
\centering
\includegraphics[width=0.32\columnwidth]{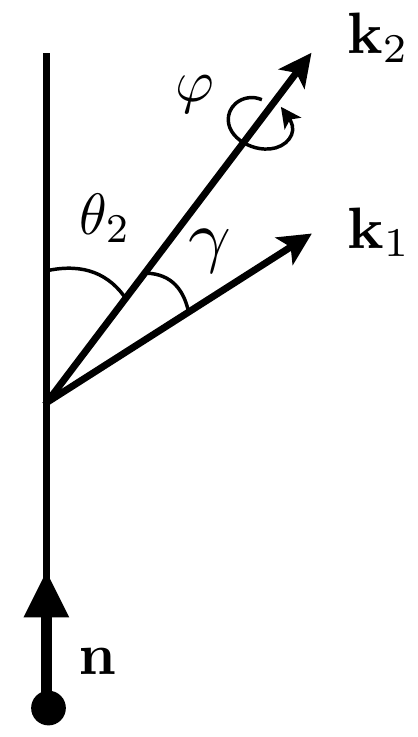}
\caption{\label{fig:angles} Definition of the angles $\theta_2, \gamma$ and $\varphi$ used in the calculation of the multipoles $Q_n$.}
\end{figure}

\subsection{Multipoles due to the intrinsic contribution}
\label{sec:mult_int}

We start by calculating the multipoles generated by the intrinsic kernel~\eqref{eq:W2int}. Let us denote by $(\theta_1, \phi_1)$ and  $(\theta_2, \phi_2)$ the angular coordinates of $\bk_1$ and $\bk_2$. Since we have fixed the direction of observation $\bn$ on the $\hat{\mathbf{z}}$ axis, we have $\widehat\bk_2\cdot\bn=\cos\theta_2$. We first do a change of variables from 
$\big\{\theta_1, \phi_1, \theta_2, \phi_2\big\}\rightarrow\big\{\gamma, \varphi, \theta_2, \phi_2\big\}$, where $\cos\gamma=\widehat\bk_1\cdot\widehat\bk_2$ and $\varphi$ is the azimutal angle of $\bk_1$ around $\bk_2$, see Fig.~\ref{fig:angles}. The Jacobian of this transformation is 1, since it is a rotation. In Eq.~\eqref{QellP}, the only quantities that depend on $\varphi$ and $\phi_2$ are the Legendre polynomials. We have
\begin{eqnarray}
\widehat{\boldsymbol{\kappa}}_{1}\!\cdot\!\mathbf{n} & \!\!= \!\!& \frac{-k_{1}\sin\gamma\sin\theta_{2}\cos\left(\varphi\!-\!\phi_{2}\right)+\left(k_{1}\!\cos\gamma\!-\!k_{2}\right)\cos\theta_{2}}{\sqrt{k_1^2+k_2^2-2k_1k_2\cos\gamma}}\,,\nonumber\\
\widehat{\boldsymbol{\kappa}}_{2}\!\cdot\!\mathbf{n} & \!\!= \!\!& \frac{-k_{1}\sin\gamma\sin\theta_{2}\cos\left(\varphi\!-\!\phi_{2}\right)+\left(k_{1}\!\cos\gamma\!+\!2k_{2}\right)\cos\theta_{2}}{\sqrt{k_1^2+4k_2^2+4k_1k_2\cos\gamma}}\,,\nonumber\\
\widehat{\boldsymbol{\kappa}}_{3}\!\cdot\!\mathbf{n} & \!\!=\! \!& \frac{2k_{1}\sin\gamma\sin\theta_{2}\cos\left(\varphi\!-\!\phi_{2}\right)-\left(2k_{1}\!\cos\gamma\!+\!k_{2}\right)\cos\theta_{2}}{\sqrt{4k_1^2+k_2^2+4k_1k_2\cos\gamma}}\,.\nonumber
\end{eqnarray}
For any value of $n$, the integral over $\phi_2$ and $\varphi$ can be done analytically, since the Legendre polynomials can always be expressed as a series of cosines. For odd $n$'s we find that the integrals vanish, as expected due to the symmetry of the line correlation function. We present here the derivation and explicit expression for the monopole $n=0$, the quadrupole $n=2$ and the hexadecapole $n=4$. In Appendix~\ref{app:multipoles}, we derive a general expression valid for any $n$.

\begin{figure*}[t]
\centering
\includegraphics[height=5.2cm]{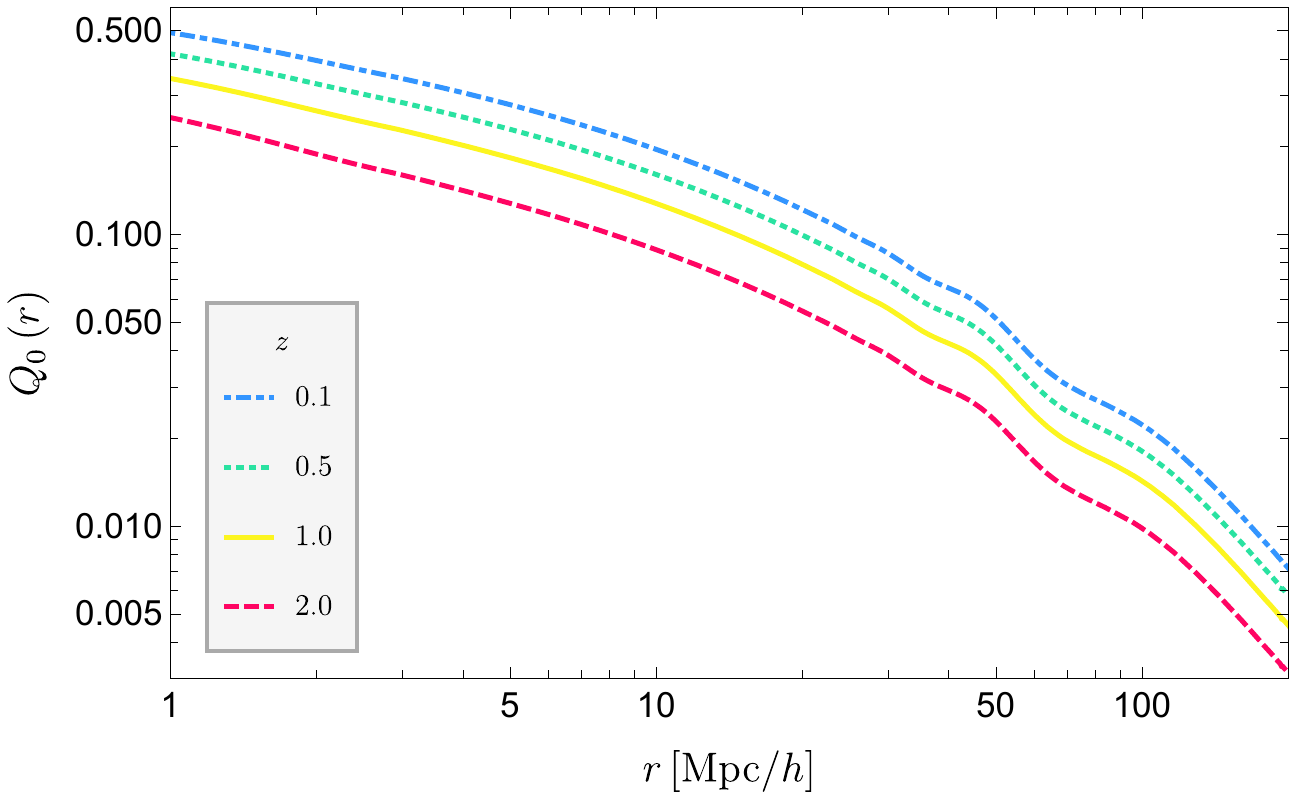}\hspace{0.2cm}\includegraphics[height=5.2cm]{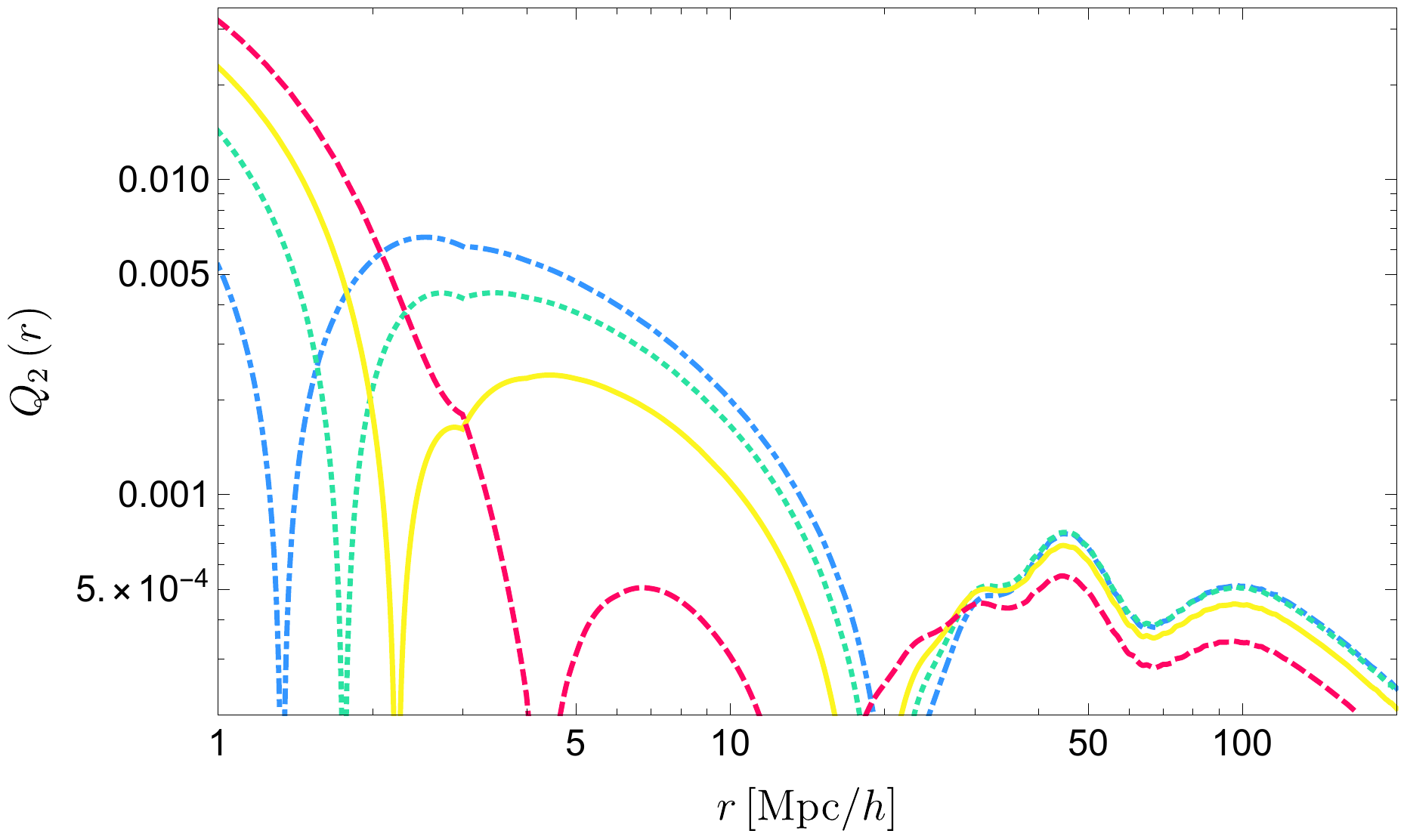}\\
\includegraphics[height=5.2cm]{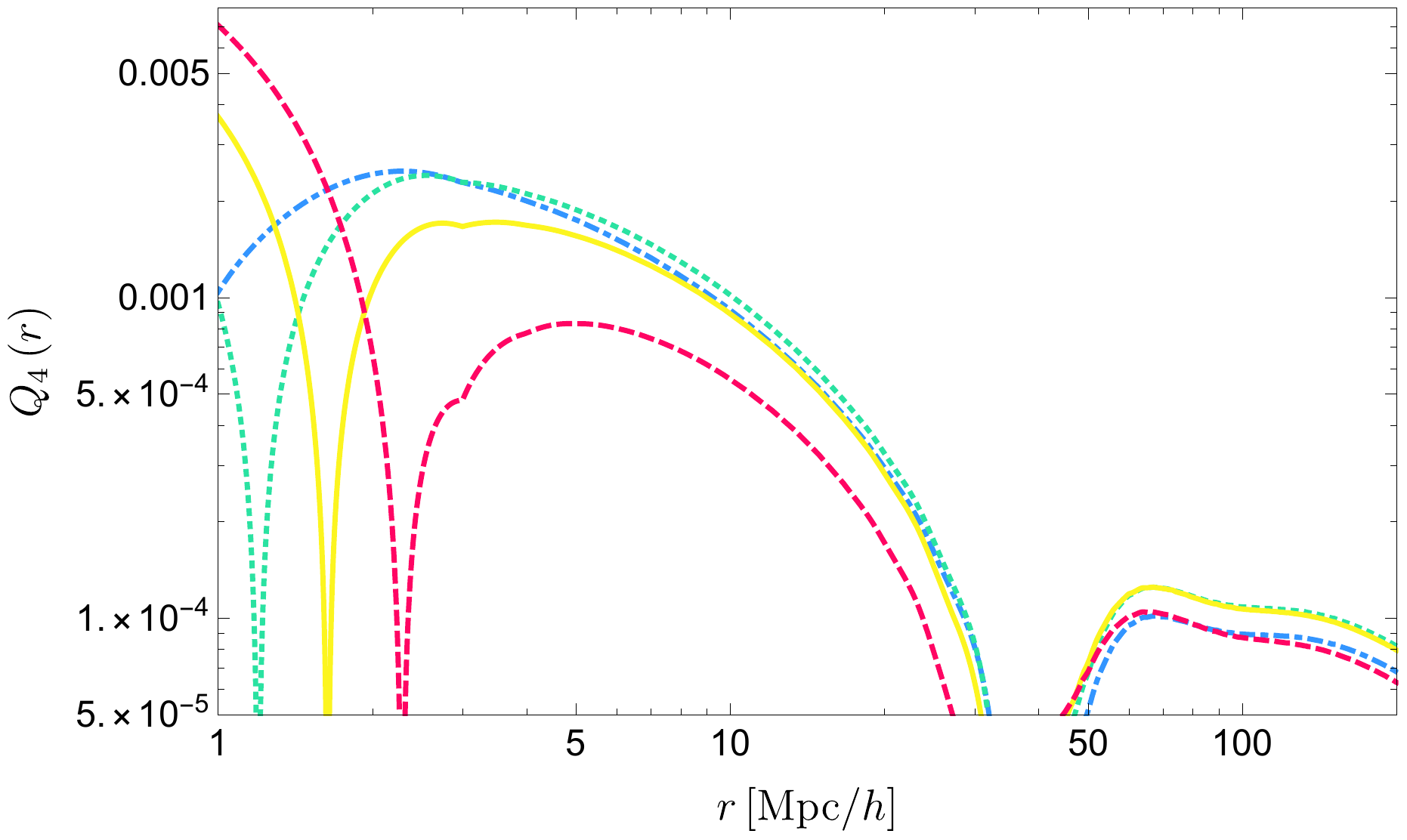}\hspace{0.4cm}\includegraphics[height=5.2cm]{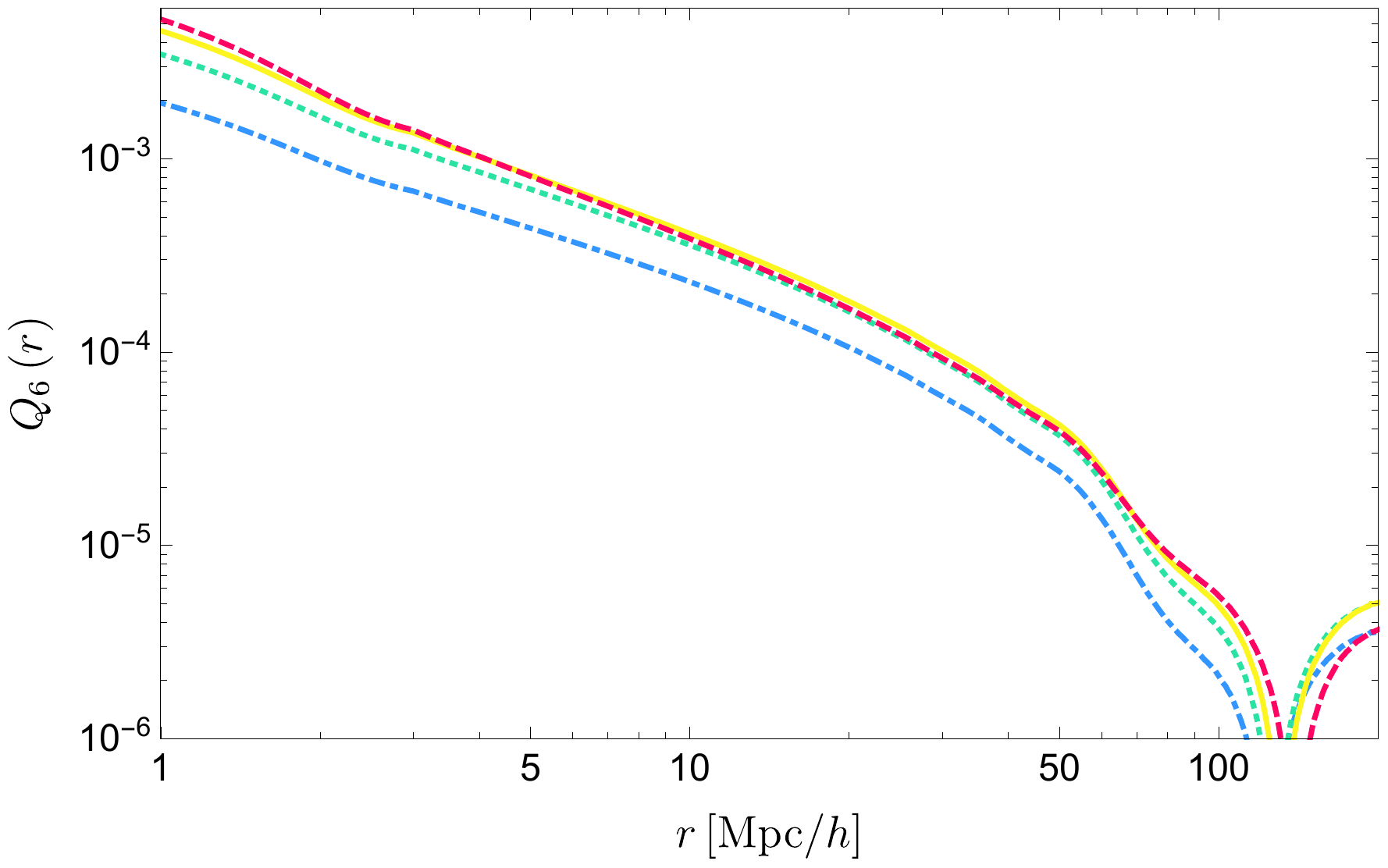}
\caption{\label{fig:mult} Multipoles of the line correlation function, plotted as a function of separation $r$, for redshift $z=0.1$ to $z=2$. The monopole is positive at all separations. The quadrupole, hexadecapole, and tetrahexadecapole change sign.}
\end{figure*}

\subsubsection{The monopole of the intrinsic contribution}

For the monopole, the integral over $\phi_2$ and $\varphi$ in Eq.~\eqref{QellP} trivially gives $4\pi^2$ since the Legendre polynomials are constant. The integral over $\theta_2$ can then be performed analytically 
\begin{align}
&\int_{-1}^1 d\mu_2\frac{b_1F_{2}+b_2/2+b_{s^2}S_2/2+\mu_2^2f\,G_{2}}{b_1+\mu_2^2f}\\
&=2\left[G_{2}+\left(F_{2}-G_{2}+\frac{b_2}{2b_1}+\frac{b_{s^2}}{2b_1}S_2\right)\frac{\arctan\sqrt{\beta}}{\sqrt{\beta}}\right]\nonumber
\end{align}
where $\mu_2=\cos\theta_2$ and 
\be
\beta\equiv\frac{f}{b_1}\, .
\ee
The monopole then simply becomes
\begin{align}
&Q^{\rm int}_0(r, z)=\frac{r^{9/2}}{8\pi\sqrt{2}}\int_{0}^{2\pi/r}\hspace{-0.2cm}dk_{1}k_{1}^{2}\int_{0}^{2\pi/r}\hspace{-0.2cm}dk_{2}k_{2}^{2}\int_{-1}^{\nu_{\text{cut}}}\hspace{-0.1cm}d\nu\nonumber\\
&\sqrt{\frac{P_{L}\left(\left|\mathbf{k}_{1}+\mathbf{k}_{2}\right|,z\right)P_{L}\left(k_{1},z\right)}{P_{L}\left(k_{2},z\right)}}\Bigg\{F_{2}(-\bk_1-\bk_2,\bk_1)\nonumber\\
&+\left(\frac{b_2}{2b_1}+\frac{b_{s^2}}{2b_1}S_2(-\bk_1-\bk_2,\bk_1)\right)\frac{\arctan\sqrt{\beta}}{\sqrt{\beta}} \nonumber\\
&+\left(F_{2}-G_2\right)(-\bk_1-\bk_2,\bk_1)\left(\frac{\arctan\sqrt{\beta}}{\sqrt{\beta}}-1\right)\Bigg\}\nonumber\\
&\quad\quad\times\sum_{i=1}^3j_0(\kappa_i r)\, ,\label{Qmono}
\end{align}
where 
\be
\nu_{\rm cut}=\min\Big\{1,\max\{-1,\big[(2\pi/r)^2-k_1^2-k_2^2\big]/[2k_1k_2]\}\Big\}\nonumber
\ee
enforces the condition $|\bk_1 + \bk_2| \leq 2\pi/r$. Here, the kernel $F_2$ and $G_2$, and the $\kappa_i$ defined in Eqs.~\eqref{kappa1} to~\eqref{kappa3}  can be expressed as functions of $k_1$, $k_2$ and $\nu=\cos\gamma=\hat\bk_1\cdot\hat\bk_2$ only. Equation~\eqref{Qmono} contains three integrals that can be computed numerically.

\subsubsection{The quadrupole of the intrinsic conctribution}

To calculate the quadrupole, we first need to integrate the terms in the square bracket in Eq.~\eqref{QellP} over $\phi_2$ and $\varphi$. As an example, let us look at the first term. We have
\begin{align}
\int_0^{2\pi}\!\!d\phi_2\int_0^{2\pi}\!\!d\varphi\ L_2(\widehat{\boldsymbol{\kappa}}_{1}\cdot\mathbf{n})=& \left(2\pi\right)^{2}\left[1-\frac{3}{2}\frac{k_1^2(1-\nu^2)}{\kappa^{2}_{1}}\right]\nonumber\\
&\times L_{2}\left(\cos \theta_{2}\right)\, . \label{intphi2}
\end{align}
As for the monopole, the integral over $\theta_2$ can then be performed analytically

\begin{align}
&\int_{-1}^{1}d\mu_{2}\frac{b_1F_{2}+b_2/2+b_{s^2}S_2/2+\mu_{2}^{2}f\,G_{2}}{b_1+\mu_{2}^{2}f}L_{2}\left(\mu_{2}\right)\nonumber\\
&=-\left(F_{2}-G_{2}+\frac{b_2}{2b_1}+\frac{b_{s^2}}{2b_1}S_2\right)\nonumber\\
&\quad\times \beta^{-3/2}\left[-3\sqrt{\beta}+\left(\beta+3\right)\arctan\sqrt{\beta}\right]\, .
\end{align}
Similar expressions can be found for the second and third terms in the square bracket of~\eqref{QellP}. Putting everything together, we then obtain for the quadrupole
\begin{align}
&Q^{\rm int}_2(r, z)=\frac{r^{9/2}}{8\pi\sqrt{2}}\int_{0}^{2\pi/r}\hspace{-0.2cm}dk_{1}k_{1}^{2}\int_{0}^{2\pi/r}\hspace{-0.2cm}dk_{2}k_{2}^{2}\int_{-1}^{\nu_{\text{cut}}}\hspace{-0.2cm}d\nu\nonumber\\
&\sqrt{\frac{P_{L}\left(\left|\mathbf{k}_{1}+\mathbf{k}_{2}\right|,z\right)P_{L}\left(k_{1},z\right)}{P_{L}\left(k_{2},z\right)}}\sum_{i=1}^3j_2(\kappa_i r)\left(1-\frac{3}{2}\frac{\rho_i^2}{\kappa_i^2} \right)\nonumber\\
&\times\left(F_{2}-G_2+\frac{b_2}{2b_1}+\frac{b_{s^2}}{2b_1}S_2\right)(-\bk_1-\bk_2,\bk_1)\nonumber\\
&\times\frac{5}{2\beta^{3/2}}\Big(-3\sqrt{\beta}+(\beta+3)\arctan\sqrt{\beta} \Big)\, ,\label{Qquad}
\end{align}
where 
\be
\rho_1^2=\rho_2^2=k_1^2(1-\nu^2)\quad \mbox{and}\quad \rho_3^2=4\rho_1^2\,.\label{rhoi}
\ee
Equation~\eqref{Qquad} contains again three integrals that can be computed numerically.

\subsubsection{The hexadecapole of the intrinsic contribution}

The hexadecapole can be calculated in a very similar way as the quadrupole. The only difference is that the integral over $\phi_2$ and $\varphi$ in Eq.~\eqref{intphi2} contains Legendre polynomial of degree four instead of two. The resulting integral over $\mu_2$ can again been done analytically and we find
\begin{align}
&Q^{\rm int}_4(r, z)=\frac{r^{9/2}}{8\pi\sqrt{2}}\int_{0}^{2\pi/r}\hspace{-0.2cm}dk_{1}k_{1}^{2}\int_{0}^{2\pi/r}\hspace{-0.2cm}dk_{2}k_{2}^{2}\int_{-1}^{\nu_{\text{cut}}}\hspace{-0.2cm}d\nu\nonumber\\
&\sqrt{\frac{P_{L}\left(\left|\mathbf{k}_{1}+\mathbf{k}_{2}\right|,z\right)P_{L}\left(k_{1},z\right)}{P_{L}\left(k_{2},z\right)}}\nonumber\\
&\times \left(F_{2}-G_2+\frac{b_2}{2b_1}+\frac{b_{s^2}}{2b_1}S_2\right)(-\bk_1-\bk_2,\bk_1)\nonumber\\
&\frac{9}{8\beta^{5/2}}\left[(3\beta^{2}+30\beta+35)\arctan\sqrt{\beta}-\left(\frac{55}{3}\beta+35\right)\sqrt{\beta} \right]\nonumber\\
&\quad\times\sum_{i=1}^3j_4(\kappa_i r)\left(1-5\frac{\rho_i^2}{\kappa_i^2} +\frac{35}{8}\frac{\rho_i^4}{\kappa_i^4} \right)\, .\label{Qhexa}
\end{align}

\subsubsection{General expression for the multipole $Q_n$ of the intrinsic contribution}

Following the same steps as for the monopole, quadrupole, and hexadecapole, one can derive a general expression for the multipole of order $n$. The detail of the derivation is presented in Appendix~\ref{app:multipoles}. Here, we only give the final expression
\begin{align} 
Q^{\rm int}_{2n}&(r,z)=\frac{r^{9/2}}{8\pi\sqrt{2}}\int_{0}^{2\pi/r}\hspace{-0.2cm}dk_{1}k_{1}^{2}\int_{0}^{2\pi/r}\hspace{-0.2cm}dk_{2}k_{2}^{2}\int_{-1}^{\nu_{\text{cut}}}\hspace{-0.2cm}d\nu\nonumber\\
&\sqrt{\frac{P_{L}\left(\left|\mathbf{k}_{1}+\mathbf{k}_{2}\right|,z\right)P_{L}\left(k_{1},z\right)}{P_{L}\left(k_{2},z\right)}}\nonumber\\
&\frac{4n+1}{2}i^{2n}\mathcal{L}_{2n}(I)\sum_{i=1}^{3}j_{2n}\left(\kappa_{i}r\right)\psi_{2n}\left(\frac{\rho_{i}}{\kappa_{i}}\right)\, .\label{eq:Q2n}
\end{align}
Here,
\begin{equation}
\mathcal{L}_{2n}\left(I\right)=2^{2n}\sum_{m=0}^{n}\begin{pmatrix}2n\\2m\end{pmatrix}
\begin{pmatrix}n+m-\frac{1}{2}\\2n\end{pmatrix}I_{2m}\,,
\end{equation}
with
\begin{align}
I_{2m} &= \frac{2}{2m+1}\Bigg\{G_{2}(-\bk_1-\bk_2,\bk_1)\\
&+\left(F_{2}-G_2+\frac{b_2}{2b_1}+\frac{b_{s^2}}{2b_1}S_2\right)(-\bk_1-\bk_2,\bk_1)\nonumber\\
&\times{}_{2}\mathcal{F}_{1}\left(1,\frac{1}{2}+m,\frac{3}{2}+m;-\beta\right)\Bigg\}\,,\nonumber
\end{align}
where $_{2}\mathcal{F}_{1}$ denotes the Gauss hypergeometric function and
\begin{equation}
\psi_{2n}\!\left(\frac{\rho_{i}}{\kappa_{i}}\right)=\,_{2}\mathcal{F}_{1}\left(-n,n+\frac{1}{2},1;\left(\frac{\rho_{i}}{\kappa_{i}}\right)^{2}\right)\,,
\end{equation}
with the $\rho_i$ defined in Eq.~\eqref{rhoi}.

\subsection{Multipoles due to the mapping contribution}

We now calculate the multipoles generated by the mapping kernel $W_2^{\rm map}$ in Eq.~\eqref{eq:W2map}. We perform the same change of variables as in Section~\ref{sec:mult_int}: $\big\{\theta_1, \phi_1, \theta_2, \phi_2\big\}\rightarrow\big\{\gamma, \varphi, \theta_2, \phi_2\big\}$ and we integrate analytically over $\varphi, \theta_2$ and $\phi_2$. All the contributions in Eq.~\eqref{eq:W2map}  can be written as a sum of the following integrals:
\begin{align}
\mathcal{I}_{nmm^{\prime}} & =\frac{1}{2\left(2\pi\right)^{2}}\int_{0}^{\pi}d\theta_{2}\sin\theta_{2}\int_{0}^{2\pi}d\phi_{2}\int_{0}^{2\pi}d\varphi\nonumber\\
&\qquad\qquad\qquad\times\frac{\mu_{1}^{m}\mu_{2}^{m^{\prime}}}{1+\beta\mu_{2}^{2}}L_{n}\left(\widehat{\boldsymbol{\kappa}}_{i}\cdot\mathbf{n}\right)\,,
\end{align}
These integrals can be performed analytically. Their expression is given in Appendix~\ref{app:multipolesapp}. The multipoles contain then three remaining integrals over $k_1$, $k_2$ and $\nu$ that can be performed numerically. As an example, here we write the expression for the monopole, but similar expressions can be derived for any multipole,
\begin{eqnarray}
Q_0^{\rm map}(r,z)&= & \frac{r^{9/2}}{8\pi\sqrt{2}}\int_{0}^{2\pi/r}\hspace{-0.2cm}dk_{1}k_{1}^{2}\int_{0}^{2\pi/r}\hspace{-0.2cm}dk_{2}k_{2}^{2}\int_{-1}^{\nu_{\text{cut}}}\hspace{-0.2cm}d\nu\nonumber\\
& &\sqrt{\frac{P_{L}\left(\left|\mathbf{k}_{1}+\mathbf{k}_{2}\right|,z\right)P_{L}\left(k_{1},z\right)}{P_{L}\left(k_{2},z\right)}}\sum_{i=1}^3j_0(\kappa_i r) \nonumber\\
& & \Bigg\{\frac{f}{2}\left(\frac{k_{2}}{k_{1}}-\frac{k_{1}k_{2}}{k_{12}^{2}}\right)\mathcal{I}_{0\,1\,1}-\frac{f}{2}\frac{k_{2}^{2}}{k_{12}^{2}}\mathcal{I}_{0\,0\,2}\nonumber\\
& &+\frac{\beta f}{2}\frac{k_{2}^{2}}{k_{12}^{2}}\mathcal{I}_{0\,2\,2}+\frac{\beta f}{2}\frac{k_{2}^{3}}{k_{1}k_{12}^{2}}\mathcal{I}_{0\,1\,3}\Bigg\}\,,\label{eq:Q0map}
\end{eqnarray}
where $k_{12}\equiv\left|\mathbf{k}_{1}+\mathbf{k}_{2}\right|$.

\section{Results}
\label{sec:result}

\begin{figure}[t]
\centering
\includegraphics[width=0.9\columnwidth]{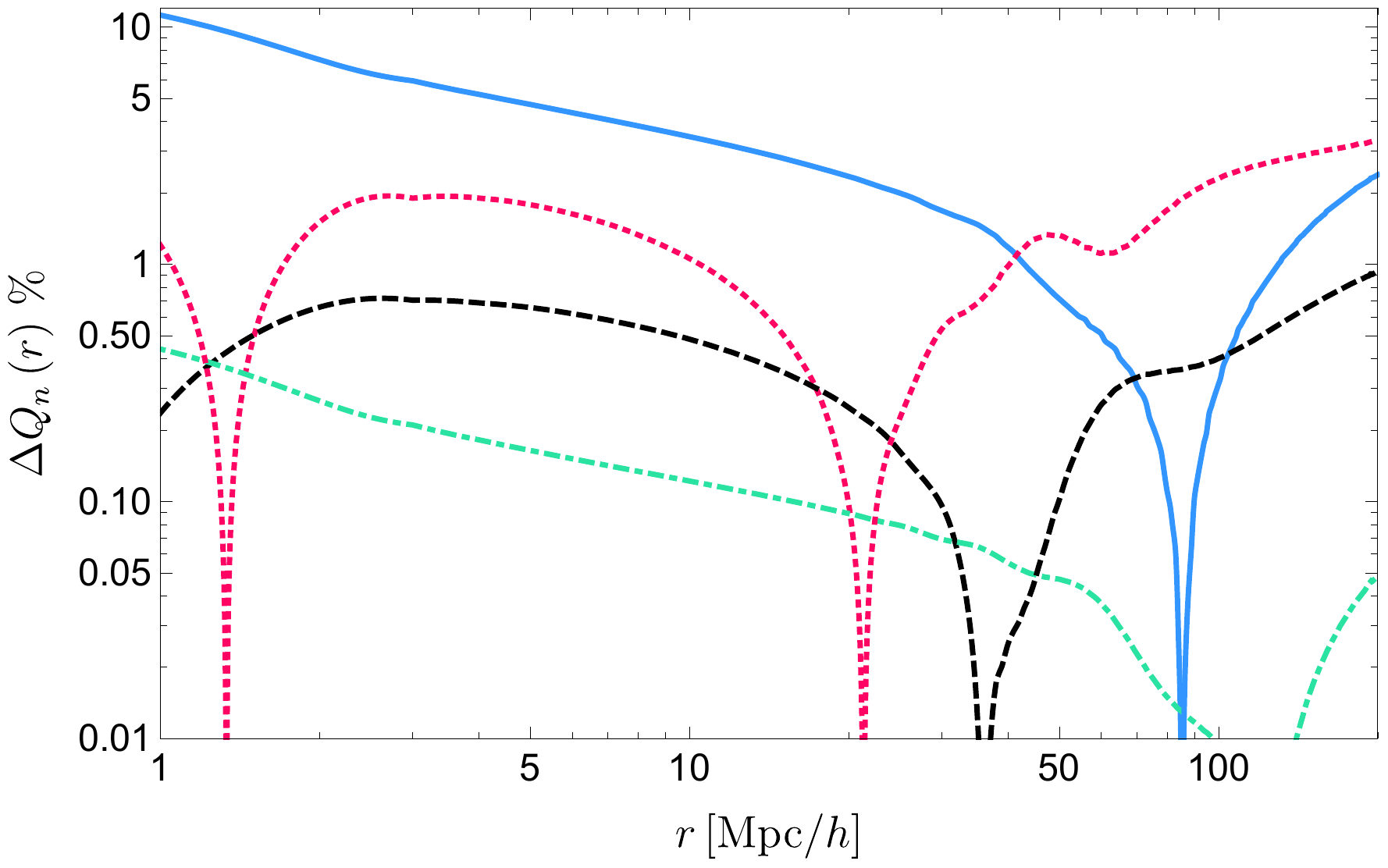}\\
\includegraphics[width=0.9\columnwidth]{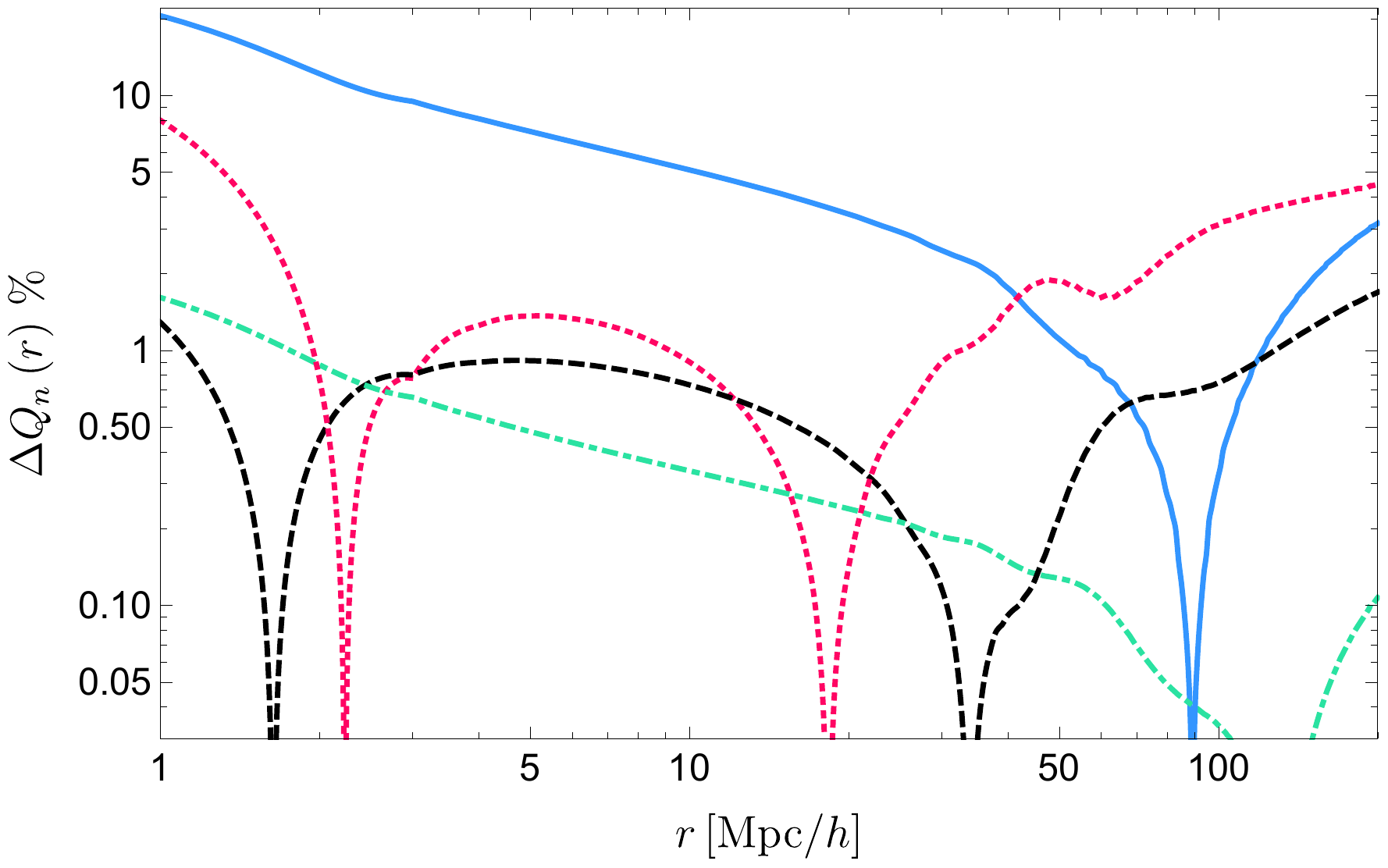}
\caption{\label{fig:relative} Relative contribution due to redshift-space distortions~\eqref{DeltaQ}, plotted as a function of separation $r$, at $z=0.1$ (top panel) and $z=1$ (bottom panel). The blue solid line shows the monopole contribution, the red dotted line the quadrupole contribution, the black dashed line the hexadecapole contribution, and the green dot-dashed line the tetrahexadecapole contribution. }
\end{figure}

\begin{figure*}[t]
	\centering
	\includegraphics[height=5.2cm]{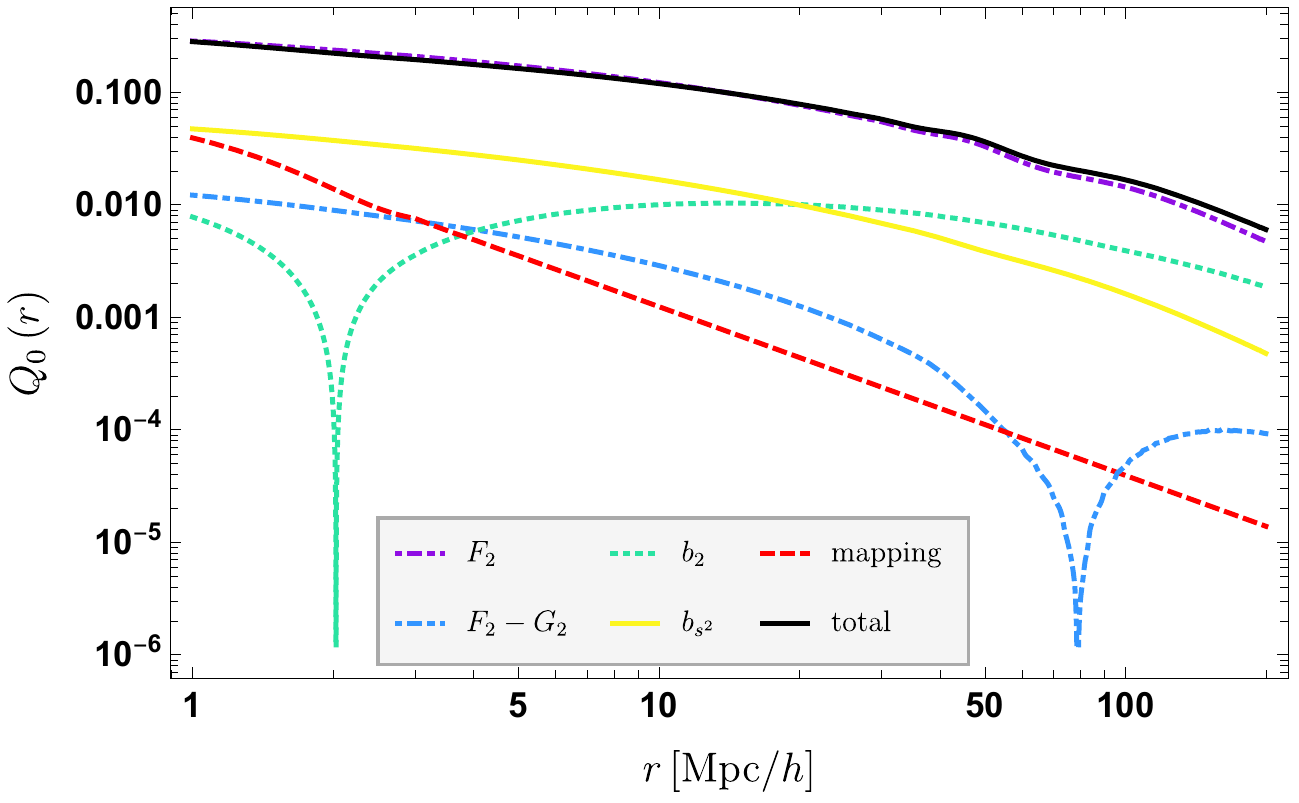}\hspace{0.2cm}\includegraphics[height=5.2cm]{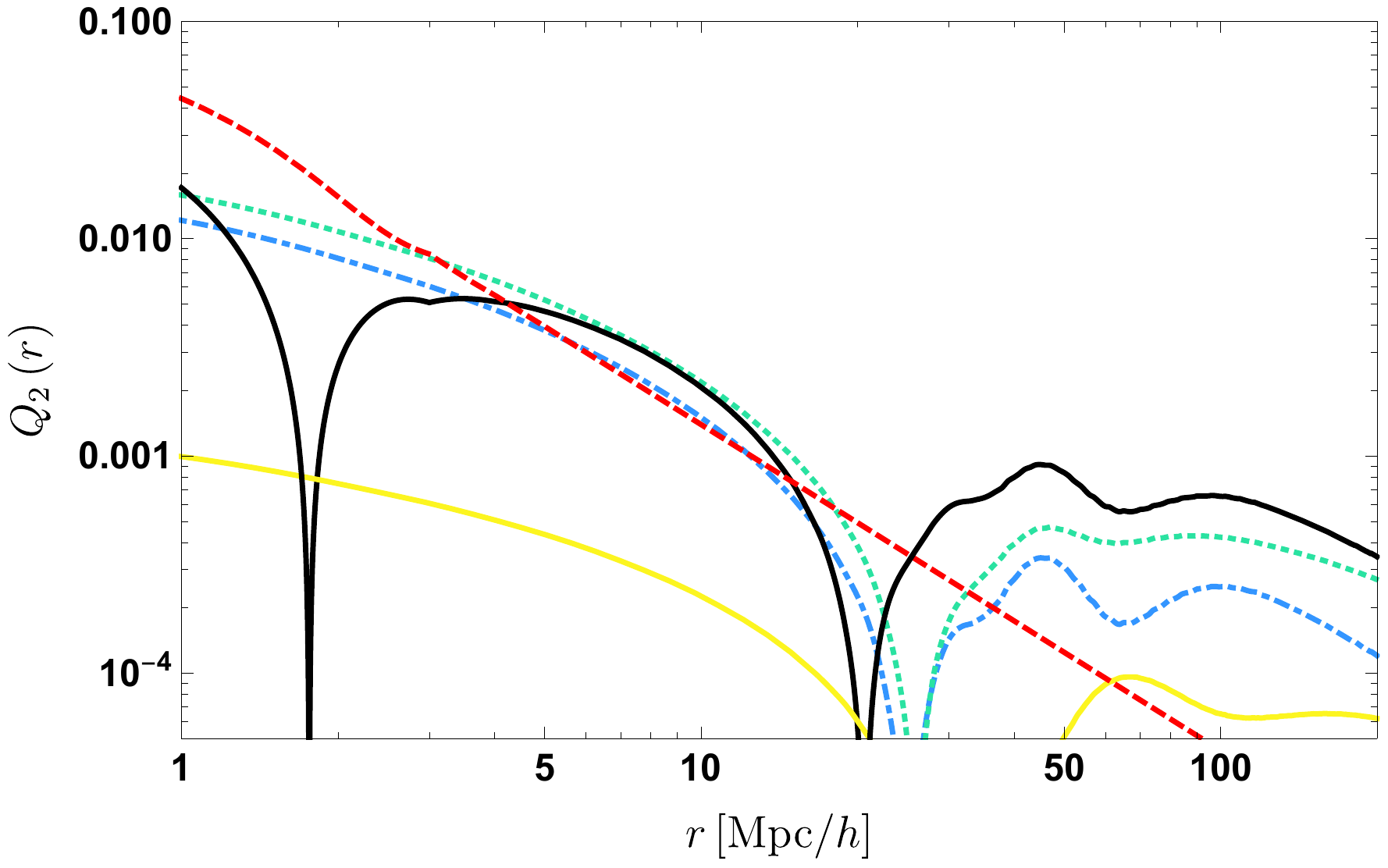}\\
	\includegraphics[height=5.2cm]{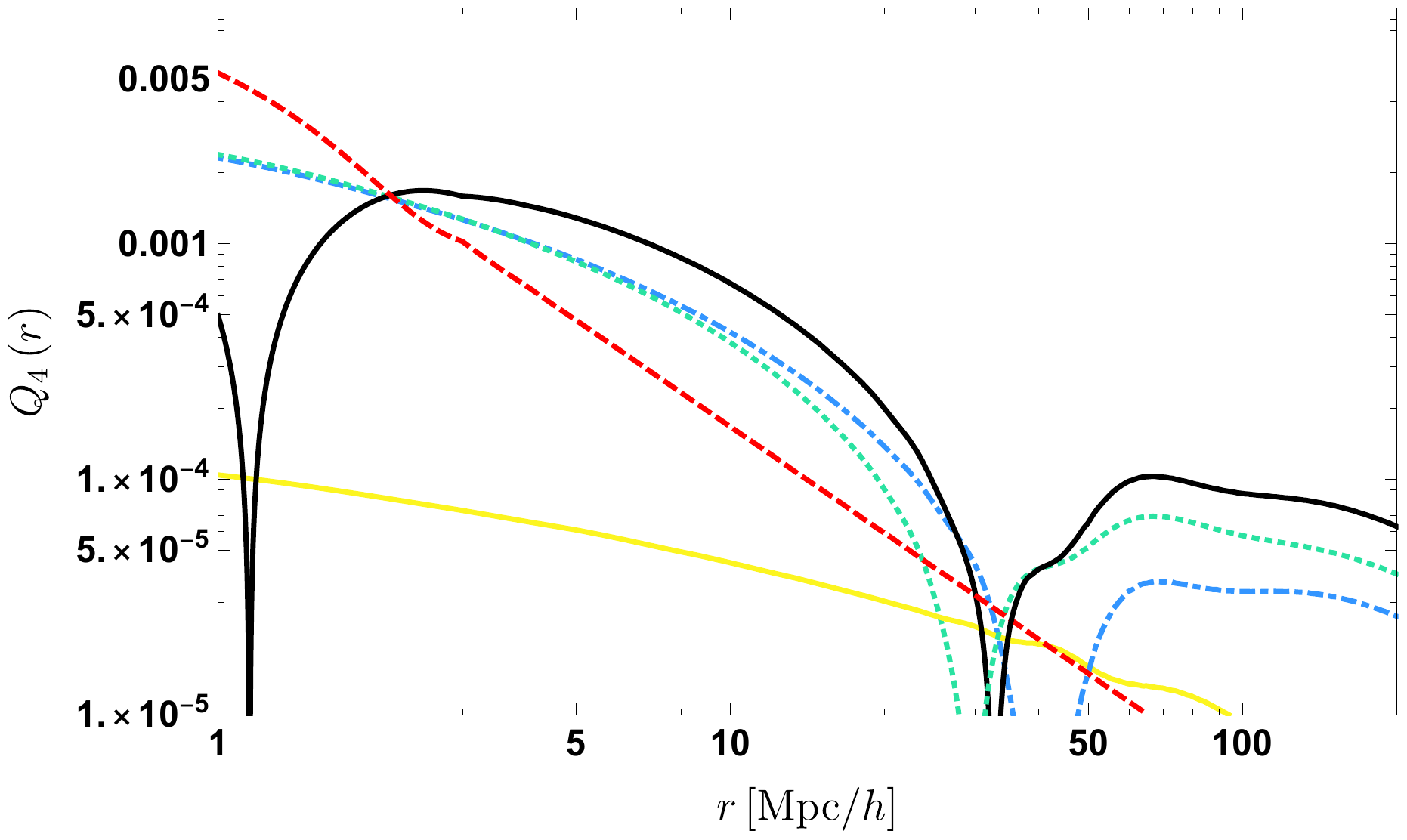}\hspace{0.4cm}\includegraphics[height=5.2cm]{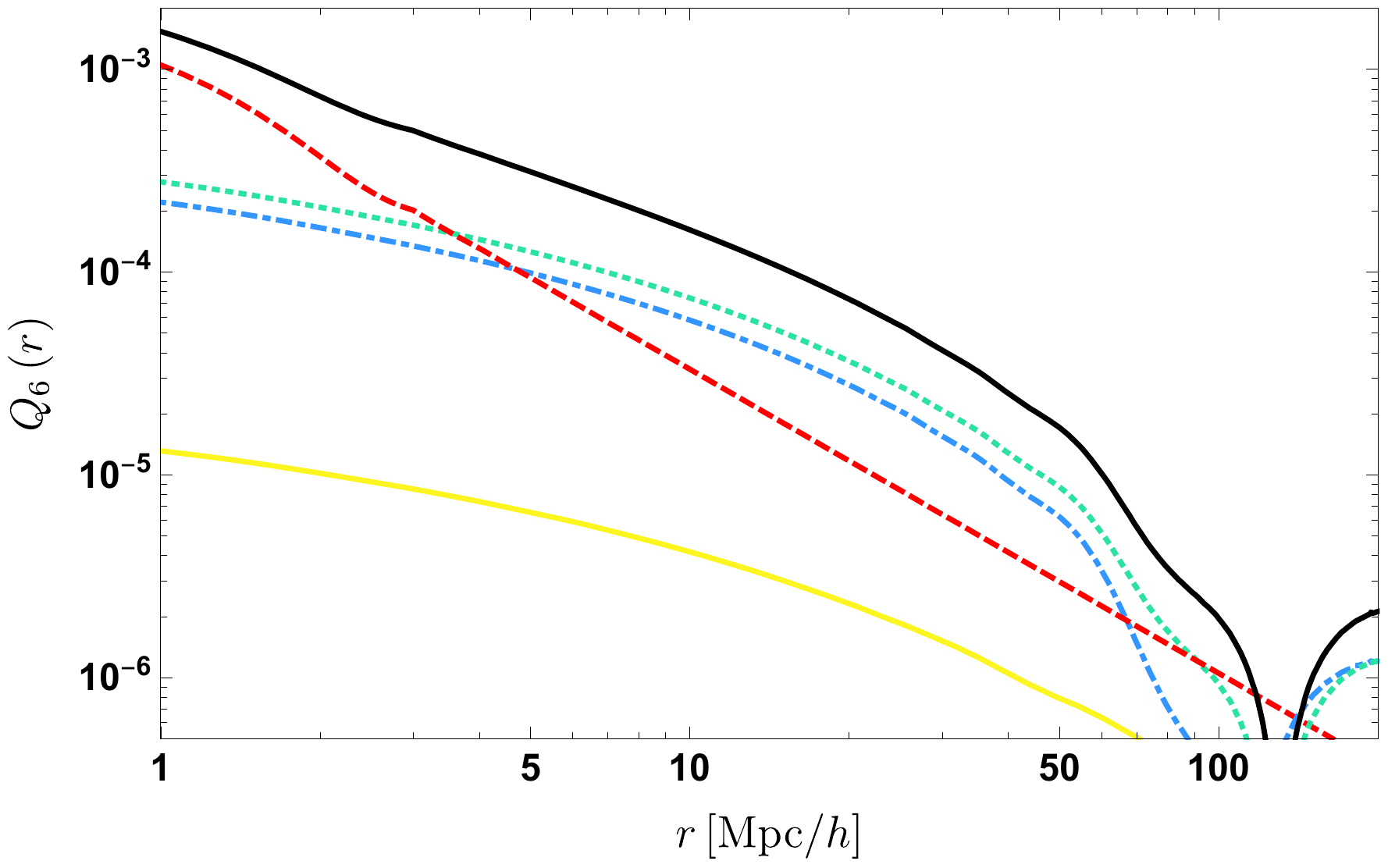}
	\caption{\label{fig:nonlinear} Different contributions to the multipoles of the line correlation function at redshift $z=1.0$. We plotted the density contribution $F_{2}$ (purple dot-dashed line), redshift-space distortion $F_{2}-G_{2}$ (blue dot-dashed line) with linear bias $b_1=2.1$, the nonlinear bias contribution (green dotted line) with $b_2=0.9$, the nonlocal bias contribution (yellow solid line) with $b_{s^2}=-4/7(b_1-1)=-0.63$, the mapping contribution (red dashed line) and the total (black solid line).}
\end{figure*}

We now calculate explicitly the multipoles of the line correlation function in a $\Lambda$CDM universe with parameters~\cite{Ade:2015xua}: $\Omega_m=0.3089, \Omega_b h^2=0.0223, h=0.6774, n_s=0.9667, \sigma_8=0.8159$. We choose as fiducial value for the bias $b_1=1, b_2=0$ and $b_{s^2}=0$, except for Fig.~\ref{fig:nonlinear} where we show explicitly the impact of nonlinear bias on the multipoles. In Fig.~\ref{fig:mult} we show the monopole, quadrupole, hexadecapole, and tetrahexadecapole ($n=6$) at different redshifts. These multipoles are the sum of the intrinsic contribution and the mapping contribution. The monopole decreases with redshift, whereas the quadrupole, hexadecapole, and tetrahexadecapole have a more compli- cated behavior. The redshift dependence is governed by the coupling kernels $F_2$ and $G_2$, by the linear power spectrum, and by the growth rate $f$, which enters in a different way in the different multipoles.

We see that the monopole dominates over the other multipoles by at least one order of magnitude. Note that the monopole is always positive, whereas the quadrupole, hexadecapole, and tetrahexadecapole change sign. The $r$-dependence of the multipole $n$ is governed by the sum of the spherical Bessel functions $j_n(\kappa_i r)$, weighted by different $k$ and $\nu$-dependent prefactors. It is therefore not surprising that the multipoles can change sign. This behavior is not specific to the line correlation function: the monopole of the two-point correlation function in redshift space does indeed also change sign at large separation; see, e.g.,\ \cite{Samushia:2013yga}. Note that, as mentioned before, our results are obtained using second-order perturbation theory. At small scales, Fingers of God are expected to generate non-negligible corrections to the line correlation function and to change, consequently, its behavior.

In Fig.~\ref{fig:relative}, we show the relative contribution due to redshift-space distortions
\be
\label{DeltaQ}
\Delta Q_n=\frac{Q_n-Q^{\rm no\, rsd}_n}{Q^{\rm no\, rsd}_0}\, ,
\ee
for $n=0,2,4, 6$ and 8. Note that $Q^{\rm no\, rsd}_0$ is equivalent to Eq.~(30) for $\ell(r, z)$ in~\cite{Wolstenhulme:2014cla}. We see that redshift-space distortions generate a correction of up to 20\% in the monopole, at small separation and high redshift. The quadrupole is roughly 5\% of the monopole and the hexadecapole one\%. The tetrahexadecapole is always less than 0.5\% of the monopole, apart at high redshift and very small separation where it reaches 2\%. Most of the information about redshift-space distortions is therefore captured by the first three even multipoles.

In Fig.~\ref{fig:nonlinear}, we compare the different contributions to the multipoles at $z=1$. We choose the value $b_1=2.1$ for the linear bias, as measured in~\cite{Gil-Marin:2014sta}, and we split the intrinsic contribution into: the redshift-space distortion part generated by the $F_2$ and $G_2$ kernels (blue line), the nonlinear bias contribution with $b_2=0.9$ (green line)~\cite{Gil-Marin:2014sta} and the nonlocal bias contribution with $b_{s^2}=-4/7(b_1-1)=-0.63$ (yellow line)~\cite{McDonald:2009dh}. The mapping contribution is shown in red, and for the monopole, we also show separately the density contribution (purple line). As expected, the main contribution to the monopole is due to the density contribution $F_{2}$ which is intrinsically isotropic. The nonlocal bias contribution has a shape very similar to the density contribution, whereas the nonlinear bias contribution is more relevant at large separation. This is consistent with the results presented in Fig. 1 of~\cite{Eggemeier:2016asq} for the monopole. Comparing the intrinsic redshift-space distortion contribution (blue line) with the mapping contribution (red line), we see that the former dominates at large scales, whereas the latter is more important at small scales. This behavior is even more pronounced for the higher multipoles ($n=2,4$ and 6), where we see that the mapping contribution is important mainly below 5\,Mpc/$h$. For these higher multipoles, contrary to the monopole, the nonlocal bias contribution is strongly subdominant. This indicates that this contribution is mainly isotropic. The nonlinear bias contribution, on the other hand, contributes to all multipoles, in a very similar way as the intrinsic redshift-space distortion contribution.

Comparing the amplitude of the density contribution in the monopole (purple line) with the intrinsic redshift-space distortions contribution in all multipoles (blue line), we see that the latter is significantly suppressed with respect to the former one. This can be understood by looking at the form
of Eqs.~\eqref{Qmono},~\eqref{Qquad} and~\eqref{Qhexa}, where we see that the intrinsic
redshift-space distortions contribution is always proportional to the {\it difference} between the density kernel $F_2$ and the velocity kernel $G_2$. This follows from the fact that the correlation between phases in Eq.~\eqref{eps3} is proportional to the weighted bispectrum~$b_\Delta\propto B_\Delta/\sqrt{P_\Delta P_\Delta P_\Delta}$. This weighted bispectrum probes the difference between the linear relation between $V$ and $\delta$ and the nonlinear relation. If these relations are the same, then $F_2=G_2$ and the function $W_2^{\rm int}$ defined in Eq.~\eqref{eq:W2int} reduces simply to $F_2$ (when $b_2=b_{s^2}=0$). We recover then the expression for the line correlation function in real space.
Hence by measuring the line correlation function in redshift space, we probe the fact that the relation between the density $\delta$ and the peculiar velocity $V$ iis different at linear and at second order in perturbation theory. In other words, we probe the difference between the continuity and Euler equation at linear and second order in perturbation theory.

As such, the line correlation function is complementary to the two-point correlation function in redshift space. The two-point correlation function probes indeed the linear relation between density and velocity by measuring the growth rate $f$. The line correlation function adds information since it probes the nonlinear relation between the density and velocity by measuring the difference $F_2-G_2$. This clearly shows that phase correlations encode a different type of information than the two-point correlation function. Modified theories of gravity generically modify both the growth rate $f$~\cite{Gleyzes:2015rua,Alonso:2016suf,Leung:2016xli} and the coupling kernels $F_2$ and $G_2$~\cite{Bernardeau:2011sf}. Hence, the line correlation function in redshift space is expected to be useful to constrain modifications of gravity.

\begin{figure}[t]
	\centering
	\includegraphics[width=0.9\columnwidth]{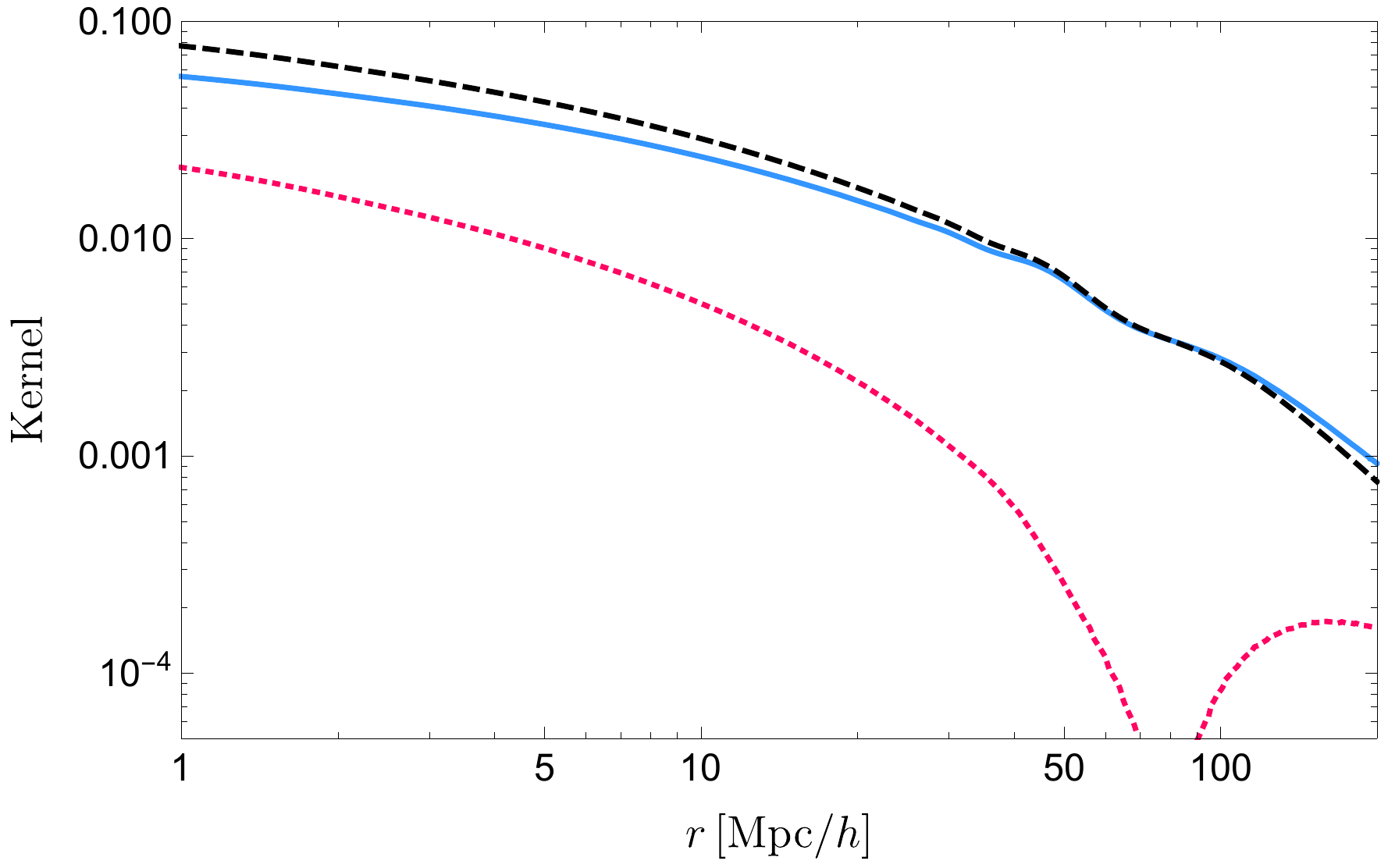}
	\caption{\label{fig:kernel} Monopole of the line correlation function~\eqref{Qmono}, at redshift $z=1$, generated by the kernel $F_2$ (blue solid line), $G_2$ (black dashed line) and the difference $F_2-G_2$ (red dotted line). The contributions from $F_2$ and $G_2$ are negative, while the difference is positive at small separations and negative at large separations.}
\end{figure}

In Fig.~\ref{fig:kernel}, we compare the contribution to the monopole~\eqref{Qmono} generated by the kernel $F_2$ only, by the kernel $G_2$ and by the difference $F_2-G_2$. We see that the difference is significantly smaller than the individual contributions from $F_2$ and $G_2$. This explains the suppression of the redshift- space distortion signal, with respect to the signal in real space. Note that here we are using second-order perturbation theory, which does not account for the effect of Fingers of God at small scales. As shown in~\cite{Gaztanaga:2005ad,Taruya:2010mx,Gil-Marin:2014pva,Hashimoto:2017klo}, those have a strong impact on the bispectrum in the nonlinear regime. In a future work, we will study the line correlation function beyond perturbation theory, accounting for the Fingers of God, to see if they enhance the multipoles.

\begin{figure}[t]
	\centering
	\includegraphics[width=0.9\columnwidth]{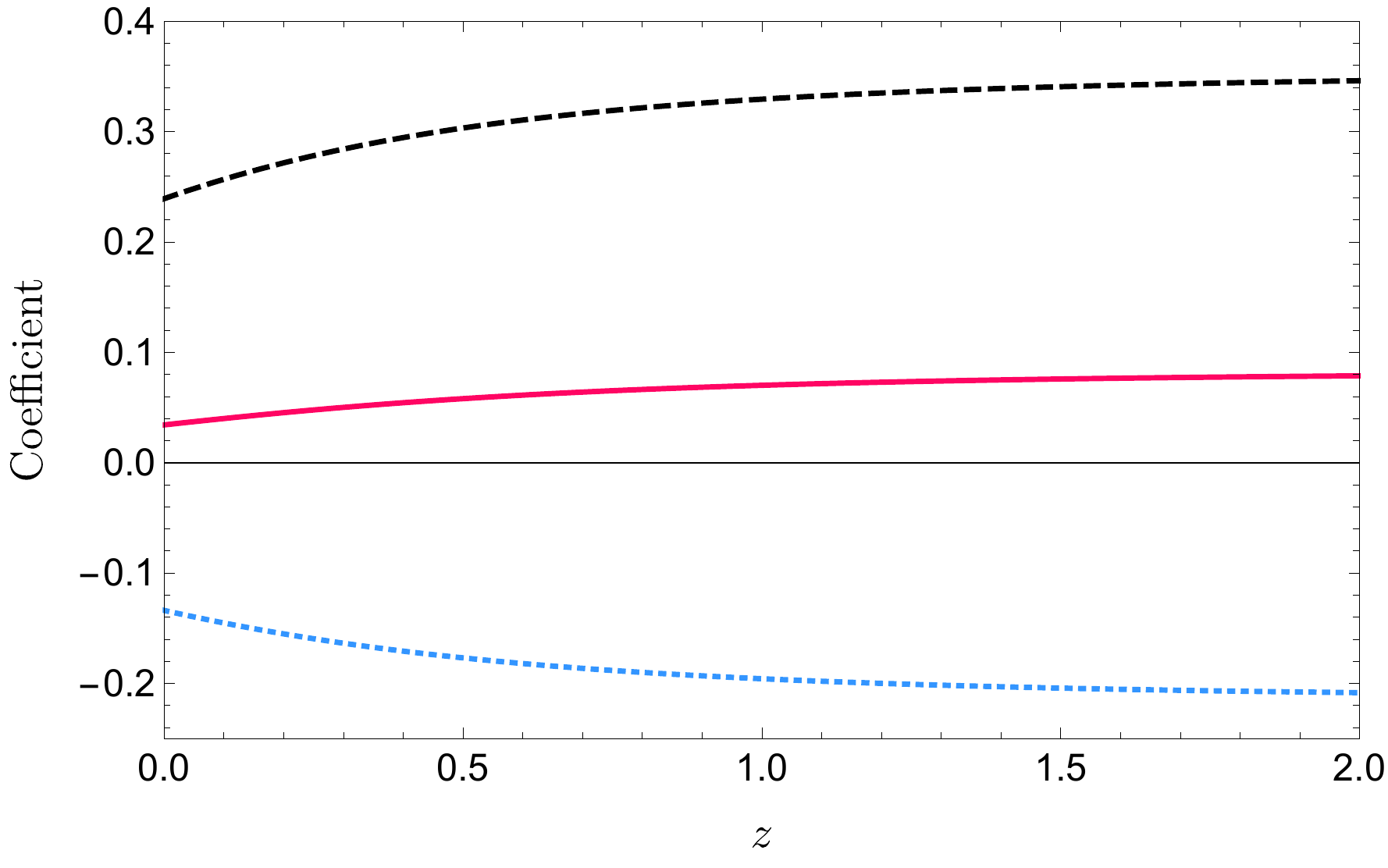}
	\caption{\label{fig:coeff} Coefficients proportional to the growth rate $\beta=f/b$, in front of the monopole~\eqref{A0} (blue dotted line), quadrupole~\eqref{A2} (black dashed line) and hexadecapole~\eqref{A4} (red solid line), plotted as a function of redshift.}
\end{figure}

In Fig.~\ref{fig:coeff}, we plot the prefactors for the intrinsic monopole, quadrupole, and hexadecapole, which depend on the growth rate $\beta=f/b$, 
\bea
A_0&=&\frac{\arctan\sqrt{\beta}}{\sqrt{\beta}}-1\,, \label{A0}\\
A_2&=&\frac{5}{2\beta^{3/2}}\Big[-3\sqrt{\beta}+(\beta+3)\arctan\sqrt{\beta} \Big]\,, \label{A2}\\
A_4&=& \frac{9}{8\beta^{5/2}}\Bigg[-\left(\frac{55\beta}{3}+35 \right)\sqrt{\beta}\nonumber\\
&&+\left(3f^2+30f+35 \right)\arctan\sqrt{\beta} \Bigg]\, .\label{A4}
\eea
We see that these prefactors evolve slowly with redshift, showing that the line correlation function is less sensitive than the two-point correlation function to variations in the growth rate. We also see that these prefactors are smaller than 1 at all redshift, which also explains why the redshift-space correction is significantly smaller than the density contribution.

Note that the different dependence of the multipoles in the growth rate is very interesting, since it provides a way of disentangling it from the parameter $\sigma_8$. The two-point correlation function measures indeed the combination $f\sigma_8$ (see e.g.~\cite{Alam:2016hwk}). The monopole of the bispectrum has been shown to measure a different combination, $f^{0.43}\sigma_8$, which in combination with the two-point function allows to disentangle $f$ and $\sigma_8$~\cite{Gil-Marin:2014sta}. Here we see, from Eqs.~\eqref{Qmono},~\eqref{Qquad} and~\eqref{Qhexa}, that the multipoles of the line correlation function are sensitive to yet three other combinations of $f$ and $\sigma_8$. Combining these measurements has therefore the potential to tighten the individual constraints on $f$ and $\sigma_8$. In a future work, we will do a detail forecast on the constraints we expect from the line correlation function on $f$, $\sigma_8$ and the coupling kernels $F_2$ and $G_2$.

\begin{figure*}[t]
\centering
\includegraphics[height=5.2cm]{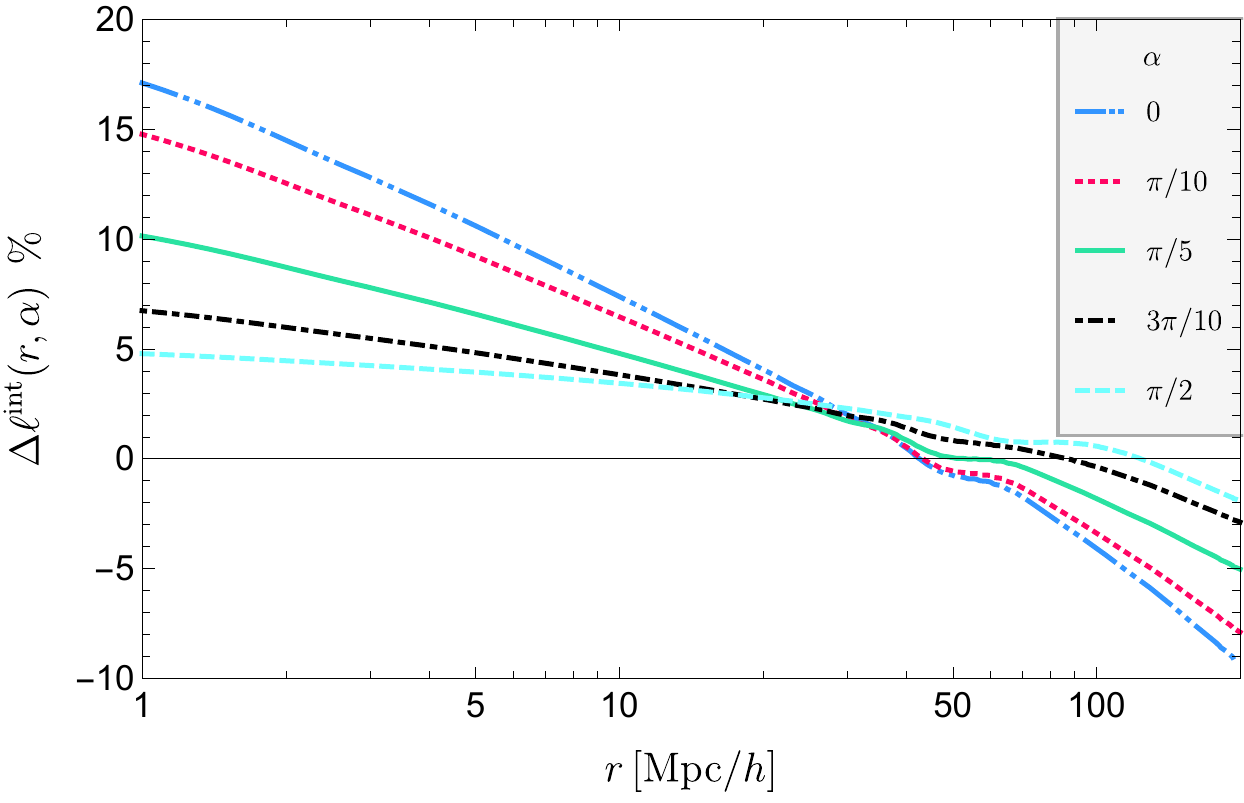}\hspace{0.2cm}\includegraphics[height=5.2cm]{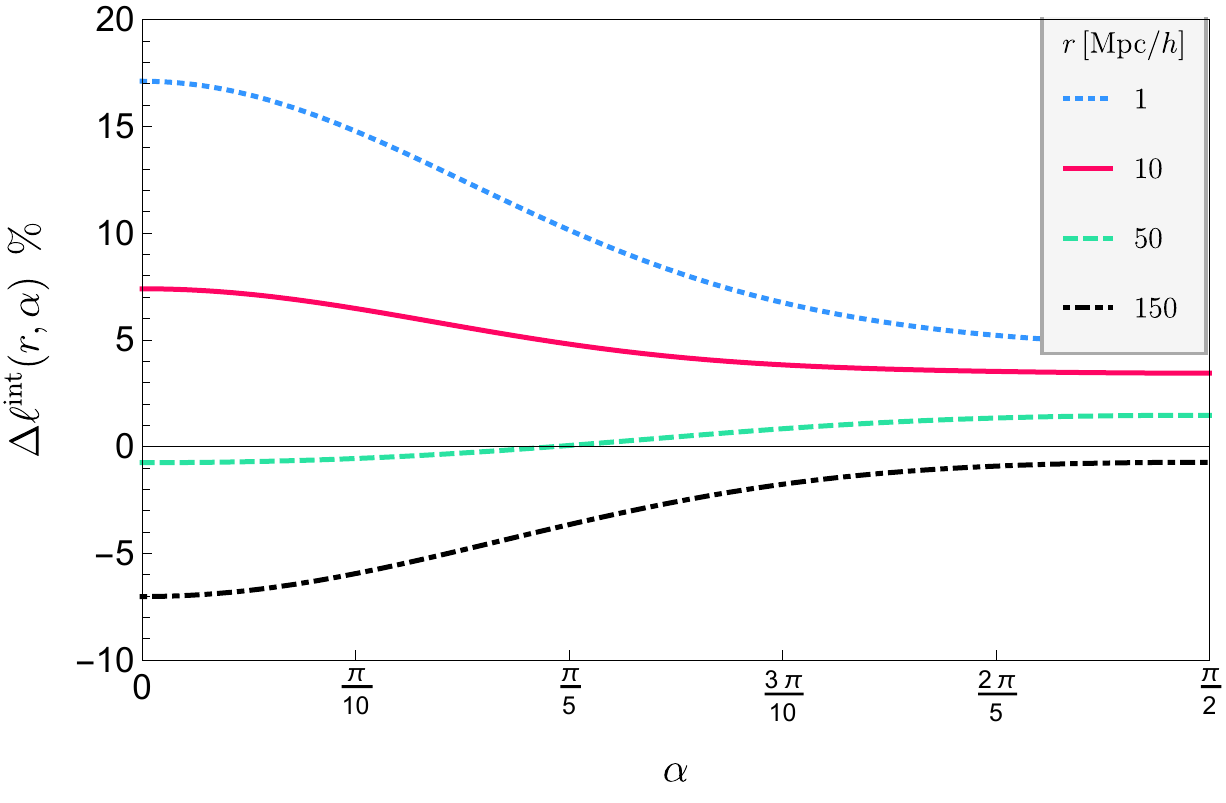}\\
\includegraphics[height=5.2cm]{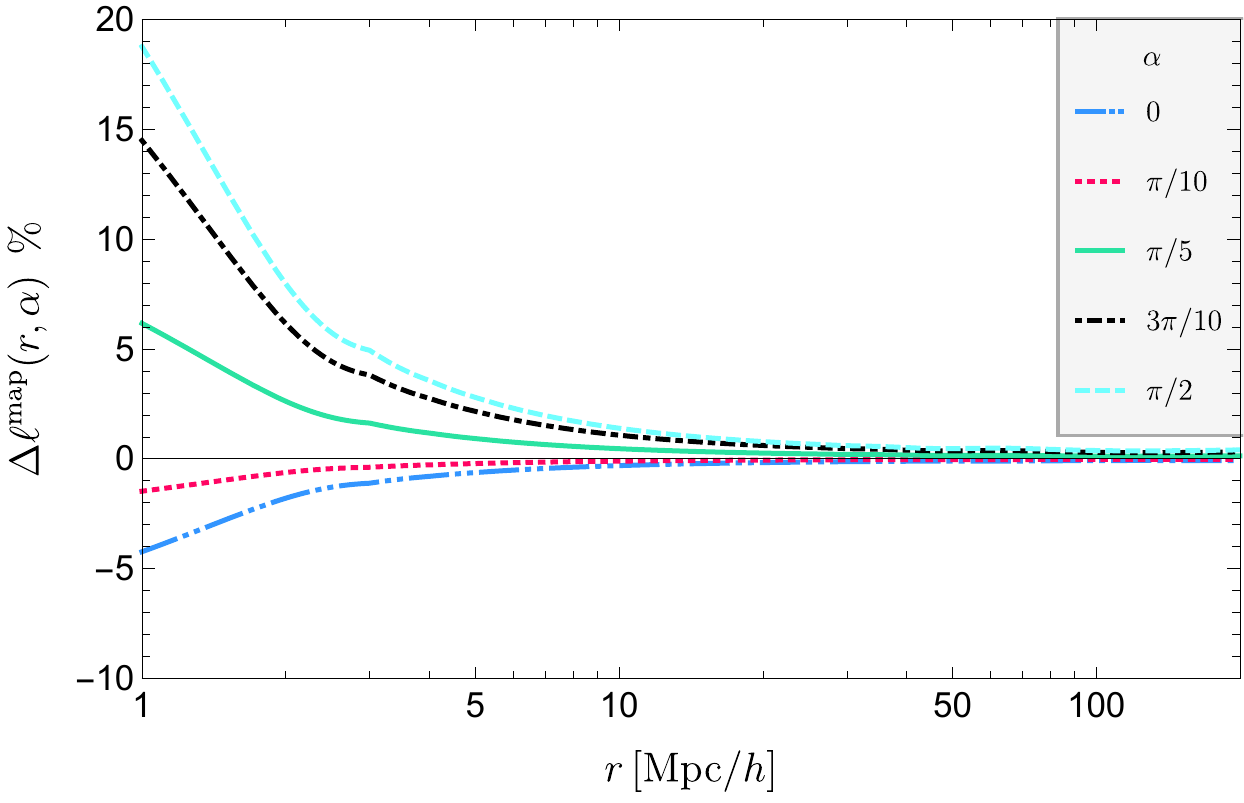}\hspace{0.4cm}\includegraphics[height=5.2cm]{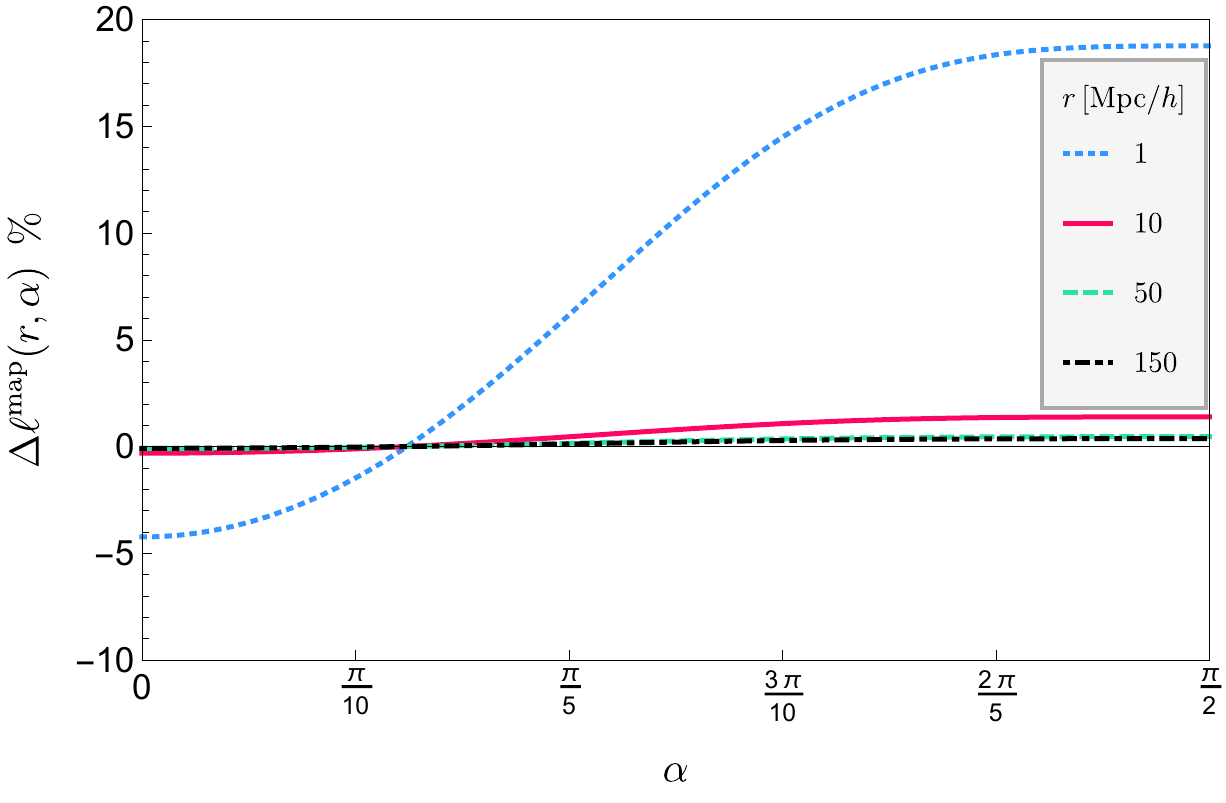}
\caption{\label{fig:relativemu} Relative contribution from redshift-space distortion~\eqref{Delta_ell}, plotted as a function of separation $r$ for fixed values of the orientation $\alpha$ (left panels); and as a function of the orientation $\alpha$ for fixed separation $r$ (right panels). The plots are at redshift $z=1$. The top panels show the intrinsic contribution, whereas the bottom panels show the mapping contribution.}
\end{figure*}

Finally, in Fig.~\ref{fig:relativemu}, we show the relative contribution from redshift-space distortion as a function of the orientation\footnote{Since the multipoles larger than $n=6$ are negligible, we can write the line correlation function as $\ell(r,\alpha,z)=Q_0(r,z)+Q_2(r,z)L_2(\cos\alpha)+Q_4(r,z)L_4(\cos\alpha)+Q_6(r,z)L_6(\cos\alpha)$.} of the line $\alpha$ and the separation $r$
\be
\label{Delta_ell}
\Delta\ell(r,\alpha, z)=\frac{\ell(r,\alpha, z)-\ell^{\rm no\, rsd}(r, z)}{\ell^{\rm no\, rsd}(r, z)}\, .
\ee
We plot separately the intrinsic contribution (top panels, with $b_1=1, b_2=0$ and $b_{s^2}=0$) and the mapping contribution (bottom panels). We see that the intrinsic contribution has the largest impact at small separation and when the three points are aligned with respect to the direction of observation ($\alpha=0$). In this case, the contribution from redshift-space distortions can reach 17\%. This reflects the fact that redshift-space distortions modify the apparent radial distance between galaxies but not their apparent angular separation. As a consequence, it is the largest when the galaxies are at different radial distances but in the same direction.

One would then naively expect that in the other extreme, i.e., when $\alpha=\pi/2$, the intrinsic contribution $\Delta\ell^{\rm int}(r,\alpha, z)$ would vanish. This corresponds indeed to the case where the three points are at the same redshift, but in different directions. From the cyan dashed line in the top left panel of Fig.~\ref{fig:relativemu} we see however that $\Delta\ell^{\rm int}(r,\pi/2, z)\neq 0$. This can be understood in the following way: suppose that the three pixels, which are at the same redshift, are all situated in an overdense region. As a consequence, the galaxies inside each pixel are falling toward the center of the pixel. The pixels in redshift space look therefore denser than they are in real space. Now since the three pixels are situated in the same overdense region, this effect induces a correlation between the three pixels. This in turns generates an additional correlation between the phases of $\Delta_{\rm int}$. This effect is independent of the orientation of $\br$ with respect to $\bn$. It simply comes from the fact that correlated density fields generate correlated velocity fields. Hence, even though at $\alpha=\pi/2$ there is no change in the apparent distance between the pixels, there is still an effect due to the fact that the size of each pixel changes in a correlated way. Note that this effect is not specific to the line correlation function, but it also exists in the two-point correlation function of galaxies: the redshift-space two-point correlation function at $\alpha=\pi/2$ is not the same as the real-space two-point correlation function $\xi(r, \alpha=\pi/2)\neq \xi^{\rm no\, rsd}(r)$.

Finally, from the bottom panels of Fig.~\ref{fig:relativemu}, we see that the mapping contribution quickly decreases with separation, for all orientations. For a fix separation, the dependance in $\alpha$ is nontrivial, going from a negative contribution when the three points are aligned with $\bn$, to a positive contribution when the three points are perpendicular to $\bn$.

\section{Conclusions}
\label{sec:conclusion}

In this paper, we have derived an expression for the line correlation function in redshift space, which is valid at second order in perturbation theory. We have expanded the line correlation function in Legendre polynomials, and we have derived a generic expression for the multipoles~$Q_n$. We have calculated explicitly the first multipoles in a $\Lambda$CDM universe, and we have found that the monopole, quadrupole, and hexadecapole encode almost all of the information in redshift space.

We have shown that the multipoles are sensitive to the difference $F_2-G_2$, i.e.\ to the difference between the nonlinear evolution of the density field and the nonlinear evolution of the velocity field. As such, the line correlation function is highly complementary to the two-point correlation function, which is sensitive to the linear growth rate of the density and velocity fields. This shows that correlations between phases encode different information than the two-point correlation function. Our expressions for the multipoles further show that each of them is sensitive to a different combination of the growth rate $f$ and of $\sigma_8$. Combining this with a measurement of the two-point correlation function, which is sensitive to the product $f\sigma_8$ can therefore break the degeneracy between these two parameters. In a future work, we will forecast how well this can be achieved with current and future surveys.

Our derivation relies on second-order perturbation theory. It is however well-known that redshift-space dis- tortions are not fully described by the second-order coupling kernel $G_2$ even on mildly nonlinear scales~\cite{Scoccimarro:1999ed,Gaztanaga:2005ad,Taruya:2010mx,Gil-Marin:2014pva,Hashimoto:2017klo}. In a future work, we will investigate how the multipoles change if we introduce nonlinear effects, like Fingers of God. We expect such effects to enhance the redshift-space distortion signal, since they will increase the difference $F_2-G_2$. 

Finally, let us note that the line correlation function targets a very particular choice of phase correlations, namely those that appear along a line, i.e., along filaments. It may be interesting to investigate other configurations, where the redshift-space distortions signal may be enhanced with respect to the real-space signal.

\section*{Acknowledgements}
We thank Pierre Fleury for useful discussions. C. B., F. O. F., and J. B. acknowledge support by the Swiss National Science Foundation. D.O. thanks for support the Research Collaboration Grant No. PG12105206 of the University of Western Australia.

\appendix

\section{Derivation of $\Delta$ at second-order}
\label{app:Delta}

The overdensity of galaxies is given by
\be
\Delta(\bn, z)=\frac{N(\bn, z)-\bar N(z)}{\bar N(z)}=\frac{\rho_g(\bn, z)\mathcal{V}(\bn, z)-\bar \rho_g(z)\bar{\mathcal{V}}(z)}{\bar \rho_g(z)\bar{\mathcal{V}}(z)}\, ,
\ee
where $N(\bn, z)$ denotes the number of galaxies detected in a pixel of volume $\mathcal{V}(\bn, z)$, situated at redshift $z$ and in direction $\bn$, and $\rho_g(\bn, z)$ is the energy density of galaxies in that pixel. Quantities with a bar refer to the average over directions, at a given redshift. At second order in perturbation theory, we obtain
\be
\Delta^{(2)}(\bn, z)=\delta_g^{(2)}(\bn, z)+\frac{\delta \mathcal{V}^{(2)}(\bn, z)}{\bar{\mathcal{V}}(z)}+\delta_g^{(1)}(\bn, z)\frac{\delta \mathcal{V}^{(1)}(\bn, z)}{\bar{\mathcal{V}}(z)}\, .
\label{eq:Deltaexpand}
\ee
Here $\delta_g^{(2)}$ and $\delta \VV^{(2)}$ are expressed in term of the observed redshift $z$. To calculate these quantities, we start from their expression written as a function of conformal time $\eta$ and we use that a fix $\eta$ is directly related to a fix background redshift $\bar z=\frac{a_0}{a}-1$, where $a_0$ denotes the scale factor today. We have
\begin{align}
&\VV(\bn,\eta)=\VV(\bn,\bar z)\simeq\bar \VV(\bar z)+\delta\VV^{(1)}(\bn,\bar z)+\delta\VV^{(2)}(\bn,\bar z)\nonumber\\
&\simeq\bar \VV(z)-\frac{\partial \bar \VV} {\partial z} \delta z+\frac{1}{2}\frac{\partial^2 \bar \VV} {\partial z^2} (\delta z)^2-\frac{\partial \delta\VV^{(1)}} {\partial z} \delta z+ \delta\VV^{(2)}\, ,\label{eq:rhopert}
\end{align}
where to obtain the second line we have used that $\bar z=z-\delta z$ and we have Taylor expanded $\VV$ around $z$ up to second-order.
We are interested in the dominant contributions to $\Delta^{(2)}$, i.e.,~those with the maximum number of radial derivatives. We can therefore neglect the second and third term in the second line of~\eqref{eq:rhopert}. For the fourth term, we rewrite the derivative with respect to $z$ as a derivative with respect to the comoving distance coordinate $\chi=\eta_O-\eta$, and we obtain at second order,
\begin{align}
\delta\VV^{(2)}&=\delta\VV^{(2)}(\bn, z)-\frac{1}{\HH}\partial_\chi \delta\VV^{(1)}\frac{\delta z}{1+\bar z}\nonumber\\
&=\delta\VV^{(2)}(\bn, z)-\frac{1}{\HH}\partial_\chi \delta\VV^{(1)} (\bv^{(1)}\cdot \bn)\, ,
\end{align}
where we have used that the dominant contribution to the redshift perturbation is due to the Doppler effect. The dominant contribution to the volume perturbation is due to redshift-space distortions. It reads
\be
\frac{\delta\VV}{\bar \VV}=-\frac{1}{\HH}\partial_\chi\left(\frac{\delta z}{1+\bar z} \right)\, .
\ee
Doing a similar expansion for $\delta \rho_g^{(2)}$ and inserting this into Eq.~\eqref{eq:Deltaexpand} we obtain
\begin{align}
\Delta^{(2)}(\bn, z)&=\delta_g^{(2)}-\frac{1}{\HH}\partial_{\chi} (\bv^{(2)}\cdot\bn)\label{eq:Deltamap}\\
&-\frac{1}{\HH}\delta_g^{(1)}\partial_{\chi} (\bv^{(1)}\!\cdot\bn)-\frac{1}{\HH}\partial_{\chi}\delta_g^{(1)}(\bv^{(1)}\!\cdot\bn)\nonumber\\
&+\frac{1}{\HH^2}\partial_{\chi}\Big[\partial_{\chi}(\bv^{(1)}\!\cdot\bn)(\bv^{(1)}\!\cdot\bn)\Big]\, . \nonumber
\end{align}
We call the first line the intrinsic contribution, since it is due to the intrinsic nonlinear evolution of the galaxy density and peculiar velocity, and the second and third line the mapping contribution, since it is generated by the nonlinear mapping between real space and redshift space. Since $\Delta$ is expressed in term of the observed redshift $z$ and the direction of the incoming photon $\bn$, we can directly Fourier transform this expression with respect to the observed coordinate $\bx_{\rm obs}=\chi(z)\bn$ to obtain Eq.~\eqref{eq:Deltamapk}.

An alternative way of deriving $\Delta(\bk)$ which is often presented in the literature~\cite{Scoccimarro:1999ed} is to do a change of variable directly in the Fourier transform. We have
\be
\Delta(\bk)=\frac{1}{(2\pi)^3}\int d^3\bx_{\rm obs}\Delta(\bx_{\rm obs})e^{-i\bk\cdot\bx_{\rm obs}}\, .\label{eq:Deltakbis}
\ee
Since the number of galaxies in a given pixel is conserved by redshift-space distortions, we can write $\left(1+\Delta(\bx_{\rm obs})\right)d^3\bx_{\rm obs}=\left(1+\delta_g(\bx)\right)d^3\bx$. Using that $d^3\bx_{\rm obs}/d^3\bx=\partial_\chi(\bv\cdot\bn)/\HH$, we obtain
\be
\Delta(\bx_{\rm obs})d^3\bx_{\rm obs}=\left(\delta_g(\bx)- \frac{1}{\HH}\partial_\chi(\bv\cdot\bn)\right)d^3\bx\, .
\ee
Inserting this into~\eqref{eq:Deltakbis} and using that $\bx_{\rm obs}=\bx+(\bv\cdot\bn)\bn/\HH$ we obtain
\begin{align}
e^{-i\bk\cdot\bx_{\rm obs}}&=e^{-i\bk\cdot\bx-i(\bv\cdot\bn)(\bk\cdot\bn)/\HH}\\
&\simeq e^{-i\bk\cdot\bx}\left(1- \frac{i}{\HH}(\bv\cdot\bn)(\bk\cdot\bn)\right)
\end{align}
we obtain
\begin{align}
&\Delta^{(2)}(\bk)=\delta_g^{(2)}(\bk)-\frac{1}{\HH}(\hat \bk\cdot\bn)^2 V^{(2)}(\bk)\label{eq:Delta2mix}\\
&-i(\bk\cdot\bn)\int \frac{d^3\bx}{(2\pi)^3}e^{-i\bk\cdot\bx}\left(\delta^{(1)}_g- \frac{\partial_\chi}{\HH}(\bv^{(1)}\cdot\bn)\right)(\bv^{(1)}\cdot\bn)\nonumber
\end{align}
The first line is the intrinsic contribution. For the second line, we Fourier transform $\delta^{(1)}$ and $\bv^{(1)}$ and we integrate over $\bx$. We obtain then the mapping contribution in Fourier space, Eq.~\eqref{eq:Deltamapk}.  We see that the two approaches give the same result in Fourier space.

\section{Calculation of the intrinsic multipoles~$Q^{\rm int}_n$}
\label{app:multipoles}

In Section \ref{sec:mult} we have performed a multipole expansion for the line correlation function. All information of our statistical measure $\ell\left(\mathbf{r},z\right)$ can be encoded in a (infinite) sum of multipoles $Q_{n}\left(r,z\right)$ given by (\ref{QellP}). Here we show how three of the six integrals in this expression can be solved analytically for any order of multipole $Q_{n}$, namely the angular integrals $\theta_{2}$, $\phi_{2}$ and $\varphi$. The multipole $Q_{n}$ contain an intrinsic and a mapping contribution; here we present the calculation for the intrinsic contribution $Q_{n}^{\text{int}}$.

Since only the kernel $W_{2}^{\text{int}}$ and the Legendre polynomial $L_{n}$ are functions of these angles, the challenge is to solve the following expression:
\begin{eqnarray}
M_{n}^{\kappa_{i}} & = & \int_{0}^{\pi}\sin\theta_{2}d\theta_{2}W_{2}^{\text{int}}\left(-\mathbf{k}_{1}-\mathbf{k}_{2},\mathbf{k}_{1},\mathbf{k}_{2},\mathbf{n}\right)\nonumber \\
&  & \int_{0}^{2\pi}d\phi_{2}\int_{0}^{2\pi}d\varphi L_{n}\left(\widehat{\boldsymbol{\kappa}}_{i}\cdot\mathbf{n}\right)\,.
\end{eqnarray}
The kernel $W_{2}^{\text{int}}$ is provided by \eqref{eq:W2map} and the vectors $\boldsymbol{\kappa}_{i}$ are defined in (\ref{kappa1})-(\ref{kappa3}). The angles $\widehat{\boldsymbol{\kappa}}_{i}\cdot\mathbf{n}$ can be written as
\begin{equation}
\widehat{\boldsymbol{\kappa}}_{i}\cdot\mathbf{n}\equiv\cos \theta_{\boldsymbol{\kappa}_{i}}=\frac{\rho_{i}\sin\theta_{2}\cos\left(\varphi-\phi_{2}\right)+\varrho_{i}\cos\theta_{2}}{\kappa_{i}}\,,\label{eq:angle}
\end{equation}
where
\begin{equation}
\begin{array}{cccccc}
\rho_{1} & = & -k_{1}\sin\gamma\, ,\quad\quad & \varrho_{1} & = & k_{1}\cos\gamma-k_{2}\, ,\\
\rho_{2} & = & -k_{1}\sin\gamma\, ,\quad\quad & \varrho_{2} & = & k_{1}\cos\gamma+2k_{2}\, ,\\
\rho_{3} & = & 2k_{1}\sin\gamma\, ,\quad\quad & \varrho_{3} & = & -2k_{1}\cos\gamma-k_{2}\, ,
\end{array}
\end{equation}
with constraint $\kappa_{i}^{2}=\rho_{i}^{2}+\varrho_{i}^{2}$. 

In order to solve the integrals, we express the Legendre polynomials as
\begin{equation}
L_{n}\left(\widehat{\boldsymbol{\kappa}}_{i}\cdot\mathbf{n}\right)=2^{n}\sum_{m=0}^{n}\begin{pmatrix}n\\
m
\end{pmatrix}\begin{pmatrix}\frac{n+m-1}{2}\\
n
\end{pmatrix}\cos^{m}\theta_{\boldsymbol{\kappa}_{i}}\,.\label{legendre}
\end{equation}
Due to the binomial coefficients, the only terms that contribute to the sum will be those with the same parity as $n$. Using the binomial expansion, $\cos^{a}\theta_{\boldsymbol{\kappa}_{i}}$ can be rewritten as
\begin{eqnarray}
\cos^{m}\theta_{\boldsymbol{\kappa}_{i}} & = & \kappa_{i}^{-m}\sum_{u=0}^{m}\begin{pmatrix}m\\
u
\end{pmatrix}\left(\varrho_{i}\cos\theta_{2}\right)^{m-u}\left(\rho_{i}\sin\theta_{2}\right)^{u}\nonumber \\
&  & \qquad\qquad\qquad\times\cos^{u}\left(\varphi-\phi_{2}\right)\, .\label{eq:coskappa}
\end{eqnarray}
Therefore, the integrals over the axial angles $\phi_2$ and $\varphi$ can be trivially solved, and it yields
\begin{align} 
\int_{0}^{2\pi}d\phi_{2}\int_{0}^{2\pi}d\varphi\cos^{u}\left(\varphi-\phi_{2}\right)&=\frac{1+\left(-1\right)^{u}}{2}\label{eq:intphi}\\
&\left(2\pi\right)^{2}\begin{pmatrix}\left(u-1\right)/2\\
-1/2
\end{pmatrix}\,.\nonumber
\end{align}
The integral over $\theta_{2}$ takes the form\footnote{To simplify the notation we drop the argument in $F_2$, $S_{s}$ and $G_2$ which are both functions of $(-\bk_1-\bk_2,\bk_1)$.}
\begin{eqnarray}
\int_{0}^{\pi}d\theta_{2}\sin\theta_{2}\frac{b\,F_{2}+b_{2}/2+b_{s^{2}}S_{2}/2+\cos^{2}(\theta_{2})\,f\,G_{2}}{b+\cos^{2}(\theta_2)\,f} & &\nonumber\\
\times\cos^{m-u}(\theta_{2})\sin^{u}(\theta_{2})\,. & &\label{eq:theta2int}
\end{eqnarray}
These two expressions above reveal an important feature about the parity of LCF. First, Eq.~\eqref{eq:intphi} tells us that only even values of $u$ contribute to the sum~\eqref{eq:coskappa}. Second, due to the orthogonality of the trigonometric functions, Eq.~\eqref{eq:theta2int} will not vanish if and only if $m-u$ is even, which implies that $m$ must also be even. Thus we conclude that in the multipole expansion (\ref{QellP}), only even multipoles contribute to the sum. This reflects the fact that the line correlation function is symmetric under the exchange of the three galaxies on the line.

Thereby, without loss of generality, one can consider a relabeling: $n\rightarrow2n$, $m\rightarrow2m$ and $u\rightarrow2u$. We can then use that $\varrho_{i}^{2\left(m-u\right)}=\left(\kappa_{i}^{2}-\rho_{i}^{2}\right)^{m-u}$, so that
\begin{equation}
\kappa_{i}^{-2m}\varrho_{i}^{2\left(m-u\right)}\rho_{i}^{2u}=\sum_{w=0}^{m-u}\begin{pmatrix}m-u\\
w
\end{pmatrix}\left(-1\right)^{w}\left(\frac{\rho_{i}}{\kappa_{i}}\right)^{2\left(u+w\right)}\,.
\end{equation}
In the interval $\theta_{2}\in\left[0,\pi\right]$ the sine function can be written as $\sin\theta_{2}=\sqrt{1-\cos^{2}(\theta_{2})}$, and consequently
\begin{equation}
\sin^{2u}(\theta_{2})=\sum_{v=0}^{u}\begin{pmatrix}u\\
v
\end{pmatrix}\left(-1\right)^{v}\cos^{2v}(\theta_{2})\,.
\end{equation}
We are now able to solve the integral over $\theta_{2}$, which is of the form,
\begin{align}
&\int_{-1}^{1}d\mu_{2}\frac{b\,F_{2}+b_{2}/2+b_{s^{2}}S_{2}/2+\mu_{2}^{2}\,f\,G_{2}}{b+\mu_{2}^{2}\,f}\mu_{2}^{2j}=\frac{2}{j+1}\qquad\nonumber\\
&\qquad\qquad\qquad\Big[G_{2}+\left(F_{2}-G_{2}+\frac{b_2}{2b_{1}}+\frac{b_{s^{2}}}{2b_{1}}S_{2}\right)\nonumber\\
&\qquad\qquad\qquad\quad\times\,_{2}\mathcal{F}_{1}\left(1,\frac{1}{2}+j,\frac{3}{2}+j;-\beta\right)\Big]\,,
\end{align}
where in our case $j=m+v-u$. Here $\,_{2}\mathcal{F}_{1}$ denotes the Gauss hypergeometric function defined by
\begin{equation}
_{2}\mathcal{F}_{1}\left(a,b,c;z\right)=\sum_{n=0}^{\infty}\frac{\left(a\right)_{n}\left(b\right)_{n}}{\left(c\right)_{n}}\frac{z^{n}}{n!}\,,
\end{equation}
where $\left(a\right)_{n}=\Gamma\left(a+n\right)/\Gamma\left(a\right)$ is the Pochhammer symbol.

Finally, collecting all these results, $M_{2n}^{\kappa_{i}}$ can be written as
\begin{align}
&M_{2n}^{\kappa_{i}}=2\left(2\pi\right)^2 2^{2n}\sum_{m=0}^{n}\sum_{u=0}^{m}\sum_{w=0}^{m-u}\sum_{v=0}^{u}
\begin{pmatrix}2n\\
2m
\end{pmatrix}
\begin{pmatrix}n+m-\frac{1}{2}\\
2n
\end{pmatrix}\nonumber\\
&\begin{pmatrix}2m\\
2u
\end{pmatrix}
\begin{pmatrix}u-\frac{1}{2}\\
-\frac{1}{2}
\end{pmatrix}\begin{pmatrix}m-u\\
w
\end{pmatrix}
\begin{pmatrix}u\\
v
\end{pmatrix}
\left(-1\right)^{w+v}\left(\frac{\rho_{i}}{\kappa_{i}}\right)^{2\left(u+w\right)}\nonumber \\
& \frac{1}{2\left(m+v-u\right)+1}\Bigg\{G_{2}(-\bk_1-\bk_2,\bk_1)\nonumber\\
&\quad\quad\quad+\Big[F_{2}-G_{2}+\frac{b_2}{2b_{1}}+\frac{b_{s^{2}}}{2b_{1}}S_{2}\Big](-\bk_1-\bk_2,\bk_1)\nonumber\\
&\times{}_{2}\mathcal{F}_{1}\left(1,\frac{1}{2}+m+v-u,\frac{3}{2}+m+v-u;-\beta\right)\Bigg\}\,.
\end{align}
This expression can be further simplified through a long algebraic manipulation and indices relabeling and we obtain
\begin{equation}
M_{2n}^{\kappa_{i}}=\left(2\pi\right)^{2}\mathcal{L}_{2n}\left(I\right)\psi_{2n}\left(\frac{\rho_{i}}{\kappa_{i}}\right)\,,
\end{equation}
where
\begin{equation}
\mathcal{L}_{2n}\left(I\right)=2^{2n}\sum_{m=0}^{n}\begin{pmatrix}2n\\
2m
\end{pmatrix}
\begin{pmatrix}n+m-\frac{1}{2}\\
2n
\end{pmatrix}I_{2m}
\end{equation}
with
\begin{align}
I_{2m} =& \frac{2}{2m+1}\Bigg\{G_{2}(-\bk_1-\bk_2,\bk_1)\nonumber\\
&+\Big[F_{2}-G_{2}+\frac{b_2}{2b_{1}}+\frac{b_{s^{2}}}{2b_{1}}S_{2}\Big](-\bk_1-\bk_2,\bk_1)\nonumber\\
&\times{}_{2}\mathcal{F}_{1}\left(1,\frac{1}{2}+m,\frac{3}{2}+m;-\beta\right)\Bigg\}\,,
\end{align}
and
\begin{equation}
\psi_{2n}\left(\frac{\rho_{i}}{\kappa_{i}}\right)=\,_{2}\mathcal{F}_{1}\left(-n,n+\frac{1}{2},1;\left(\frac{\rho_{i}}{\kappa_{i}}\right)^{2}\right)\,.
\end{equation}
This shows that the multipoles can be calculated from expression \eqref{eq:Q2n}.

\section{Calculation of the mapping multipoles~$Q^{\rm map}_n$}
\label{app:multipolesapp}

The calculation for the mapping contribution $Q_{n}^{\text{map}}$ is similar to what was done for the intrinsic contribution. Here, the expression is
\begin{eqnarray}
\mathcal{M}_{n}^{\kappa_{i}} & = & \int_{0}^{\pi}\sin\theta_{2}d\theta_{2}W_{2}^{\text{map}}\left(-\mathbf{k}_{1}-\mathbf{k}_{2},\mathbf{k}_{1},\mathbf{k}_{2},\mathbf{n}\right)\nonumber \\
&  & \int_{0}^{2\pi}d\phi_{2}\int_{0}^{2\pi}d\varphi L_{n}\left(\widehat{\boldsymbol{\kappa}}_{i}\cdot\mathbf{n}\right)\,.
\end{eqnarray}
with the kernel $W_{2}^{\text{map}}$ given by \eqref{eq:W2int}. For this case it is better to express $\mathcal{M}_{n}^{\kappa_{i}} $ as a linear combination of integrals of the type
\begin{align}
\mathcal{I}_{nmm^{\prime}} & =\frac{1}{2\left(2\pi\right)^{2}}\int_{0}^{\pi}\sin\theta_{2}d\theta_{2}\int_{0}^{2\pi}d\phi_{2}\int_{0}^{2\pi}d\varphi\nonumber\\
&\qquad\qquad\qquad\times\frac{\mu_{1}^{m}\mu_{2}^{m^{\prime}}}{1+\beta\mu_{2}^{2}}L_{n}\left(\widehat{\boldsymbol{\kappa}}_{i}\cdot\mathbf{n}\right)\,,
\end{align}
since the kernel is a sum of powers of $\mu_{1}$ and $\mu_{2}$ with the denominator $1+\beta\mu_{2}^{2}$. From \eqref{eq:W2int} it is easy to see the relation between $\mathcal{M}_{n}^{\kappa_{i}}$ and $\mathcal{I}_{nmm^{\prime}}$
\begin{eqnarray}
\mathcal{M}_{n}^{\kappa_{i}} & = & \frac{f}{2}\left(\frac{k_{2}}{k_{1}}-\frac{k_{1}k_{2}}{k_{12}^{2}}\right)\mathcal{I}_{n\,1\,1}-\frac{f}{2}\frac{k_{2}^{2}}{k_{12}^{2}}\mathcal{I}_{n\,0\,2}\nonumber\\
& &+\frac{\beta f}{2}\frac{k_{2}^{2}}{k_{12}^{2}}\mathcal{I}_{n\,2\,2}+\frac{\beta f}{2}\frac{k_{2}^{3}}{k_{1}k_{12}^{2}}\mathcal{I}_{n\,1\,3} \,.
\end{eqnarray}
with $k_{12}\equiv\left|\mathbf{k}_{1}+\mathbf{k}_{2}\right|$.
To calculate the terms $\mathcal{I}_{nmm^{\prime}}$ one can use the same tricks already used previously, namely express the Legendre polynomials as
\begin{equation}
L_{n}\left(\widehat{\boldsymbol{\kappa}}_{i}\cdot\mathbf{n}\right)=2^{n}\sum_{b=0}^{n}\begin{pmatrix}n\\
b
\end{pmatrix}\begin{pmatrix}\frac{n+b-1}{2}\\
n
\end{pmatrix}\cos^{b}\theta_{\boldsymbol{\kappa}_{i}}\,,\label{legendremap}
\end{equation}
such that
\begin{eqnarray}
\cos^{b}\theta_{\boldsymbol{\kappa}_{i}} & = & \kappa_{i}^{-b}\sum_{c=0}^{b}\begin{pmatrix}b\\
c
\end{pmatrix}\left(\varrho_{i}\cos\theta_{2}\right)^{b-c}\left(\rho_{i}\sin\theta_{2}\right)^{c}\nonumber \\
&  & \qquad\qquad\qquad\times\cos^{c}\left(\varphi-\phi_{2}\right)\, .\label{eq:coskappa}
\end{eqnarray}
 Meanwhile due the change of variables performed in Section~\ref{sec:mult_int}, one obtains the relation  $\mu_{1}\equiv\cos\theta_{1}=\cos\gamma\cos\theta_{2}-\sin\gamma\sin\theta_{2}\cos\left(\varphi-\phi_{2}\right)$. Consequently,
\begin{eqnarray}
\mu_{1}^{m} & = & \sum_{a=0}^{m}\begin{pmatrix}m\\
a
\end{pmatrix}\left(-1\right)^{a}\cos^{m-a}\gamma\sin^{a}\gamma\nonumber \\
&  & \times\cos^{m-a}\theta_{2}\sin^{a}\theta_{2}\cos^{a}\left(\varphi-\phi_{2}\right)\, .\label{eq:cos1}
\end{eqnarray}
In this way the integration over the angles $\phi_{2}$ and $\varphi$ gives
\begin{align} 
\int_{0}^{2\pi}d\phi_{2}\int_{0}^{2\pi}d\varphi\cos^{a+c}\left(\varphi-\phi_{2}\right)&=\frac{1+\left(-1\right)^{a+c}}{2}\label{eq:intphi}\\
&\left(2\pi\right)^{2}\begin{pmatrix}\frac{a+c-1}{2}\\
-1/2
\end{pmatrix}\,,\nonumber
\end{align}
implying that $a+c$ must be even; otherwise, those integrals vanish. That result allows us to rewrite the sin function as
\begin{equation}
\sin^{a+c}\theta_{2}=\left(1-\mu_{2}^{2}\right)^{\frac{a+c}{2}}=\sum_{d=0}^{\frac{a+c}{2}}\begin{pmatrix}\frac{a+c}{2}\\
d
\end{pmatrix}\left(-1\right)^{d}\mu_{2}^{2d}\,,
\end{equation}
since $\theta_{2}\in\left[0,\pi\right]$, i.e. $\sin\theta_{2}\in\left[0,1\right]$. So the integration over $\theta_{2}$ can be fully written in term of $\mu_{2}$, and it has the form
\begin{eqnarray}
\int_{-1}^{1}d\mu_{2}\frac{\mu_{2}^{X}}{1+\beta\mu_{2}^{2}} & = & \frac{1+\left(-1\right)^{X}}{1+X} \\
& & {}_{2}\mathcal{F}_{1}\left(1,\frac{1+X}{2},\frac{3+X}{2};-\beta\right)\,. \nonumber
\end{eqnarray}
Thereby, the terms $\mathcal{I}_{nmm^{\prime}}$ can be generally written as
\begin{align}
&\mathcal{I}_{nmm^{\prime}}= 2^{2n}\sum_{a=0}^{m}\sum_{b=0}^{n}\sum_{c=0}^{b}\sum_{d=0}^{\frac{a+c}{2}}
\begin{pmatrix}m\\
a
\end{pmatrix}
\begin{pmatrix}n\\
b
\end{pmatrix}
\begin{pmatrix}\frac{n+b-1}{2}\\
n
\end{pmatrix}\nonumber\\
&\qquad\qquad\begin{pmatrix}b\\
c
\end{pmatrix}
\begin{pmatrix}\frac{a+c}{2}\\
d
\end{pmatrix}
\begin{pmatrix}\frac{a+c-1}{2}\\
-1/2
\end{pmatrix}
\left(-1\right)^{a+d}\frac{1+\left(-1\right)^{a+c}}{2}\nonumber \\
&\qquad\qquad\qquad\qquad\cos^{m-a}\gamma\sin^{a}\gamma\left(\frac{\varrho_{i}}{\kappa_{i}}\right)^{b}\left(\frac{\rho_{i}}{\varrho_{i}}\right)^{c}\nonumber\\
&\qquad\quad\frac{1+\left(-1\right)^X}{2\left(1+X\right)}{}_{2}\mathcal{F}_{1}\left(1,\frac{1+X}{2},\frac{3+X}{2};-\beta\right)\,. \label{eq:Inmmprime}
\end{align}
with $X\equiv m+m^{\prime}-a+b-c+2d$. For sake of simplicity we present below only the terms that contribute to the monopole and the quadrupole,
\begin{allowdisplaybreaks}
\begin{eqnarray}
\mathcal{I}_{0\,0\,0} & = & H_{0}\,,\nonumber\\
\mathcal{I}_{0\,0\,2}& = & \frac{H_{2}}{3}\,,\nonumber\\
\mathcal{I}_{0\,1\,1} & = & \frac{H_{2}}{3}\cos\gamma\,,\nonumber\\
\mathcal{I}_{0\,2\,2} & = & \frac{H_{4}}{5}\cos^{2}\gamma+\frac{5H_{2}-3H_{4}}{30}\sin^{2}\gamma\nonumber\\
\mathcal{I}_{0\,1\,3} & = & \frac{H_{4}}{5}\cos\gamma\,,\nonumber\\
\mathcal{I}_{2\,0\,0} & = & \left[1-\frac{3}{2}\left(\frac{\rho_{i}}{\kappa_{i}}\right)^{2}\right]\frac{H_{2}-H_{0}}{2}\,,\nonumber\\
\mathcal{I}_{2\,0\,2}& = & \left[1-\frac{3}{2}\left(\frac{\rho_{i}}{\kappa_{i}}\right)^{2}\right]\frac{9H_{4}-5H_{2}}{30}\,,\nonumber\\
\mathcal{I}_{2\,1\,1} & = & \left[1-\frac{3}{2}\left(\frac{\rho_{i}}{\kappa_{i}}\right)^{2}\right]\frac{9H_{4}-5H_{2}}{30}\cos\gamma\nonumber\\
& &+\frac{\rho_{i}\varrho_{i}}{\kappa_{i}^{2}}\frac{3H_{4}-5H_{2}}{10}\sin\gamma \nonumber\\
\mathcal{I}_{2\,2\,2} & = & \left[1-\frac{3}{2}\left(\frac{\rho_{i}}{\kappa_{i}}\right)^{2}\right]\frac{15H_{6}-7H_{4}}{70}\cos^{2}\gamma\nonumber\\
& &+\Bigg\{\left[1-\frac{3}{2}\left(\frac{\rho_{i}}{\kappa_{i}}\right)^{2}\right]\frac{-H_{6}+2H_{4}-H_{2}}{8}\nonumber\\
& &+\frac{15H_{6}-42H_{4}+35H_{2}}{840}\Bigg\}\sin^{2}\gamma\nonumber\\
& &+\frac{\rho_{i}\varrho_{i}}{\kappa_{i}^{2}}\frac{15H_{6}-21H_{4}}{35}\cos\gamma\sin\gamma\,,\nonumber\\
\mathcal{I}_{2\,1\,3} & = & \left[1-\frac{3}{2}\left(\frac{\rho_{i}}{\kappa_{i}}\right)^{2}\right]\frac{15H_{6}-7H_{4}}{70}\nonumber\\
& &+\frac{\rho_{i}\varrho{i}}{\kappa_{i}^{2}}\frac{15H_{6}-21H_{4}}{70}\sin\gamma\,.\nonumber
\end{eqnarray}
where $H_{X}\equiv{}_{2}\mathcal{F}_{1}\left(1,\left(1+X\right)/2,\left(3+X\right)/2;-\beta\right)$. For higher multipoles, the expressions are more complicated but they can be straightforwardly obtained by Eq.~\eqref{eq:Inmmprime}.
\end{allowdisplaybreaks}

\bibliography{bibphase.bib}

\end{document}